\documentclass[10pt,a4paper,superscriptaddress,aps,prd,nofootinbib,notitlepage]{revtex4-1}
\usepackage[utf8]{inputenc}
\usepackage{fontenc}
\usepackage{units}
\usepackage{amsmath, amssymb, commath, mathtools}
\usepackage{physics}
\usepackage{graphicx}
\usepackage{subcaption}
\usepackage{float}
\usepackage{multirow}
\usepackage{booktabs}
\usepackage{color}
\usepackage{ulem}

\newcommand\indice[1]{_{\mathrm{#1}}}
\newcommand\Ei[1]{\mathrm{Ei}(#1)}

\newcommand\e{\mathrm{e}}
\newcommand\totalderiv[2]{\frac{\mathrm{d} #1}{\mathrm{d} #2}}

\newcommand\einheit[1]{\mathrm{#1}}
\newcommand\ten[1]{\phantom{\cdot}10^{#1}}

\begin{document}

\title{Testing gravity with the Milky Way: Yukawa potential}

\author{Jakob Henrichs}
\affiliation{Institut für Theoretische Physik, Ruprecht-Karls-Universit\"{a}t Heidelberg, Philosophenweg 16, D-69120 Heidelberg, Germany}

\author{Margherita Lembo}
\affiliation{Dipartimento di Fisica e Scienze della Terra, Universit\`a degli Studi di Ferrara, via Giuseppe Saragat 1, I-44122 Ferrara, Italy}
\affiliation{Istituto Nazionale di Fisica Nucleare, Sezione di Ferrara, via Giuseppe Saragat 1, I-44122 Ferrara, Italy}

\author{Fabio Iocco}
\affiliation{Universit\`a di Napoli ``Federico II'' \& INFN Sezione di Napoli, Complesso Universitario di Monte S.~Angelo, via Cintia, 80126 Napoli, Italy}

\author{Luca Amendola}
\affiliation{Institut für Theoretische Physik, Ruprecht-Karls-Universit\"{a}t Heidelberg, Philosophenweg 16, D-69120 Heidelberg, Germany}

%\date{\today}

\begin{abstract}

We test a Yukawa correction to the Newtonian potential, making use of our own Galaxy -- the Milky Way -- as a testbed. We include as free parameter the Yukawa strength and range and the dark matter profile parameters, and compare several morphologies for the bulge, gas, and disk components, also using Bayesian model selection criteria. 
We employ up--to--date datasets for both the visible (baryonic) component of the Milky Way, and for the tracers of the gravitational potential (the Rotation Curve).
We find that the data are consistent with the Newtonian potential, and constrain
the Yukawa coupling $\beta$ to be negative and $\lambda$ to range along the curve $\lambda=a|\beta|^c$ with $a=(0.77\pm 0.06)\,\mathrm{kpc}$ and $c=-0.503^{+0.016}_{-0.019}$.

\end{abstract}



\maketitle

%%%%%%%%%%%%%%%%%%%%%%%%%%%%%%%%%%%%%%%
%%%%%%%%%%%%%%%% Sec I %%%%%%%%%%%%%%%%
%%%%%%%%%%%%%%%%%%%%%%%%%%%%%%%%%%%%%%%
\section{Introduction}

Testing gravity is one of the most active areas of contemporary research. Although there are quite stringent constraints on deviation from Newtonian or Einsteinian gravity, they have been so far only at laboratory and solar system scales
(see e.g. \cite{Sakstein_2018,Lee_2020}). 

One of the simplest way to test gravity is to constrain the existence of a Yukawa additional term, such that the one-particle Newtonian potential assumes the form
\begin{equation}
   \Psi(r)= -\frac{GM}{r}\left(1+\beta e^{-mr}  \right)
\label{eq:modgravpot1}
\end{equation}
where $\beta$ and $m^{-1}$ express the strength of the correction and its range, respectively. Such a correction arises whenever the force is mediated by a scalar particle of mass $m$. Hence, testing gravity is equivalent to testing the existence of a fifth force of scalar nature.

The local gravity tests might be insufficient to constrain the parameters $\beta,m$ because of two reasons. Firstly, the screening phenomenon present in several modified gravity theories  \cite{amendola2010dark, Clifton_2012}, is such that the fifth force felt around massive objects is different from the one in less dense regions. Secondly, local gravity experiments do not test the gravity exerted, or experienced, by dark matter. In fact, in general, the coupling $\beta$ between two particles is the product of the Yukawa couplings $\alpha_p$ of both particles.
Let us assume for instance there are only two couplings, $\alpha_{b}$ (for baryons) and $\alpha_{dm}$ (for dark matter). Then, if a planet feels the Sun's gravity,  $\beta=\alpha_{b}^2$. If instead a star is under the action of the dark matter halo,  then $\beta=\alpha_{b} \alpha_{dm}$. Clearly,  local tests can measure $\alpha_{b}$ but not $\alpha_{dm}$.

It is important to remark that the coupling $\beta$ can be positive or negative. For instance, in a scalar-tensor theory with conformal coupling in which several matter species labelled by $i$ are coupled to gravity via  rescaled metrics $\hat g_{(i)\alpha\beta}=A_{i}^2(\phi)g_{\alpha\beta}$ one has \cite{1992CQGra...9.2093D}
\begin{equation}
    \alpha_i=\frac{d\log A_i(\phi)}{d\phi} \,.
\end{equation}
Therefore, while $\beta$  has to be positive when considering the interaction between two particles of the same species,  there is no {\it a priori} restriction on the individual values of $\alpha_p$. Thus, $\beta=\alpha_{b} \alpha_{dm}$ can be negative without the theory being necessarily unhealthy. For this reason, we include both positive and negative values of $\beta$.

To move beyond local gravity, the next target is the Milky Way. The rotation curve of the Milky Way is well known observationally, since abundant data of gas, stars and masers allow to trace the gravitational potential, under a (well posed) assumption of dynamical equilibrium, which consequently allows to constrain $\beta=\alpha_{b} \alpha_{dm}$. 
Since we test gravity well beyond the local gravity realm, we do not assume any screening for $\alpha_{dm}$ and, consequently, we leave $\beta$ as a free parameter in our analysis. 
Summarizing, our goal here is to put constraints on $\beta$ under a minimum of assumptions. In particular, we emphasize that we are not aiming at replacing  dark matter with a modified gravity.

This exercise has been performed several times in the past, using external galaxies (see e.g. \cite{1984A&A...136L..21S,2001PhRvD..63d3503D,Capozziello:2008ny,2011PhRvD..83h4038M,2011PhRvD..84l4023S, 2011MNRAS.414.1301C, 2013PhRvD..87f4002S,2013MNRAS.436.1439M,2014PhRvD..89j4011R,Almeida2018}),  and  without including dark matter (for a more detailed discussion of these earlier works, we refer to \cite{Almeida2018}). The Milky Way has been also used several times to test different modified gravity theories, see e.~g.~ \cite{Islam_2020, Dutta_2018, Moffat_2015, Moffat_2008,Islam_2019} .
In this paper, we restrict our attention to the extremely rich dataset of the Milky Way, which has been used extensively to test dark matter and alternatives to it, see e.~g.~ \cite{Negrelli:2018gll, Iocco:2015iia, Lisanti:2018qam}.
Our analysis is different from the previous ones because of various reasons: 
1) we are not necessarily trying to explain dark matter away with the Yukawa term (also because this would open problems at cosmological scales), and therefore we include both dark matter and modified gravity (as in \cite{2003PhRvL..91n1301P}); 
2) we assume that baryons are very weakly coupled to gravity, $0<|\alpha_b|\ll 1$, to pass local gravity constraints, while  we leave $\alpha_{dm}$ completely free; in other words, only dark matter is effectively coupled to gravity;
3) we use the most complete set of rotational data, including the very recent one by Eilers et al. \cite{Eilers2018}; 
4) we carefully decompose the baryonic components of the Galaxy, i.e. the bulge, the disk and the gas, and we adopt several models for each of them;
5) we exploit Bayesian criteria to gauge the relative merit of the various models. Because of the novel approach and of the larger amount of data, we believe that the constraints we derive are more robust and more general than those previously obtained.

Lastly, it is worth mentioning that the small-distance rotational data are problematic: the assumption of circular velocities probably breaks down and the observational data show significant discrepancies among different subsets. 
Cutting out data below a radial distance of 3 kpc, we conclude that the current data are well modeled by the pure Newtonian potential but, at the same time, that the constraints on the Yukawa correction are rather weak, and a non-negligible contribution is still possible. 

The structure of the paper is as follows. In Sec.~\ref{section:modified_gravity}, we introduce the underlying modified gravity theory, and in Sec.~\ref{section:components_of_the_milky_way} we describe the different baryonic components of the Milky way. Then, we calculate the contribution to the circular velocity due to the dark matter halo in Sec.~\ref{section:circular_velocity} and discuss the used data sets in Sec.~\ref{section:data}. Afterwards, the fitting procedure and the model selection criteria are discussed in Sec.~\ref{section:maximum_likelihood} and Sec.~\ref{section:model_selection}, respectively. Finally, we present our results in Sec.~\ref{section:results} and 
we finish with some conclusions in Sec.~\ref{section:conclusion}.

%%%%%%%%%%%%%%%%%%%%%%%%%%%%%%%%%%%%%%%
%%%%%%%%%%%%%%%% Sec II %%%%%%%%%%%%%%%
%%%%%%%%%%%%%%%%%%%%%%%%%%%%%%%%%%%%%%%
\section{Modified Gravity and gravitational Potential}\label{section:modified_gravity}

In this work, we will concentrate on modified gravity theories based on a single scalar field. The most general way to describe such a case is with the Horndeski theory. The corresponding Lagrangian is \cite{Amendola2019}:
\begin{equation}
S=\int d^4 x \sqrt{-g} \sum^5_{i=2} \mathcal{L}_i +S\indice{m},
\end{equation}
where $g$ describes the determinant of the metric, \(S\indice{m}\) is the action for matter and \(\mathcal{L}_i\) is given by:
\begin{align}
\mathcal{L}_{2} &= K(\phi, X)\,;\\
\mathcal{L}_3 &= - G_{3}(\phi, X) \square \phi \,; \\
\mathcal{L}_{4} &=G_{4}(\phi, X) R+G\indice{4, X}\left[(\square \phi)^{2}-\left(\nabla_{\mu} \nabla_{\nu} \phi\right)^{2}\right] \,; \\ 
\mathcal{L}_{5} &=G_{5}(\phi, X) G_{\mu \nu} \nabla^{\mu} \nabla^{\nu} \phi-\frac{G\indice{5, X}}{6}\left[(\square \phi)^{3}-3(\square \phi)\left(\nabla_{\mu} \nabla_{\nu} \phi\right)^{2}+2\left(\nabla_{\mu} \nabla_{\nu} \phi\right)^{3}\right].
\end{align}
Where \(\phi\) is the scalar field, \(X=-(g_{\mu\nu}\partial_\mu\phi\partial_\nu\phi )/2 \) is the canonical kinetic term and \(K(\phi,X)\) and \(G_3(\phi,X)\) are non-canonical kinetic functions. The functions \(G_4(\phi,X)\) and \(G_5(\phi,X)\) are coupling functions. In the weak field limit, the perturbed FLRW metric for flat space, (\(|\Omega\indice{k0}|\ll 1\)), is
\begin{equation}
ds^2=-(1+2\Psi) dt^2 + a^2(t)(1+2\Phi)\delta_{ij} dx^i dx^j \,.
\end{equation}
The potential \(\Psi\) is the Newtonian potential, which gives, as part of the time-time component of the metric, the motion to the non-relativistic particles, while the other potential \(\Phi\) influences the motion of relativistic particles \cite{Bertschinger2011}. Since the stars in a galaxy are non-relativistic objects, only the first potential is important for this work. It is possible to find a Poisson equation for the first potential. Solving this equation with a spherical matter distribution as $\rho(r)$, we get the following gravitational potential for non-relativistic objects \cite{Amendola2019}:
\begin{equation}
\Psi(r)=-h_{1}\mathrm{G} \int \frac{1}{|\mathbf{r}^\prime-\mathbf{r}|}\left(1+\beta e^{-m|\mathbf{r}^\prime-\mathbf{r}|}\right) \rho(r^\prime) d^3\mathbf{r}^\prime \,,
\label{eq:modgravpot}
\end{equation}
i.e., as anticipated in Eq.~\eqref{eq:modgravpot1}, a Newtonian potential with a Yukawa correction, where $\beta,m$ are in general function of the field $\phi$. We use the parameter $\lambda\equiv 1/m$ as a scaling length for the correction.

From now on we assume that the configuration of the field $\phi$ in a galaxy halo is approximately constant, so that $\beta,m$ are constant in the regime investigated here.

%%%%%%%%%%%%%%%%%%%%%%%%%%%%%%%%%%%%%%%
%%%%%%%%%%%%%%% Sec III %%%%%%%%%%%%%%%
%%%%%%%%%%%%%%%%%%%%%%%%%%%%%%%%%%%%%%%
\section{Components of the Milky Way}\label{section:components_of_the_milky_way}

Our galaxy, the Milky Way, is usually modelled as the sum of an extended stellar disk, a less extended central stellar ``bulge'', a (gravitationally barely relevant) disk of gas, all embedded in a dark matter halo.

The stellar and gas (hereafter, ``baryonic'') components have been carefully studied in the literature, yet no full agreement has been reached within the community over the exact distribution of both the bulge and disk. It is for this reason that rather than choosing only one description for each of the components, we prefer to explore several competing models, in order to bracket the effect of this systematic ignorance over our final results. We anticipate that -- consistently with similar analysis -- the detail of baryonic distribution does not affect our bulk conclusions, although it has a non-negligible quantitative impact on the results.

 In the following we describe in turn the different baryonic components we adopt.
 
\subsection{Bulge}
For the Milky Way the bulge can be modelled as a bar lying with its long axis in the galactic plane \cite{McMillan2018}.  We use two different models to describe the bulge. In the first one, denoted  as G2, the bulge density profile is described by a Gaussian shape, \cite{Novati2008}: 
\begin{equation}
\rho\indice{G2}(x, y, z)=\rho\indice{0,G 2} \exp \left(-0.5\phantom{\cdot}r\indice{G 2}^{2}\right),
\end{equation}
with the  radius
\begin{equation}
r\indice{G 2}=\left\{\left[\left(\frac{x}{x\indice{0,G 2}}\right)^{2}+\left(\frac{y}{y\indice{0,G 2}}\right)^{2}\right]^{2}+\left(\frac{z}{z\indice{0,G 2}}\right)^{4}\right\}^{1 / 4}.
\end{equation}
For further analysis, we transform the mass profile into spherical coordinates. Moreover, we take into account the orientation of the bulge. It can be distinguished between an in-plane rotation and an out-of-plane tilt around the angle $\beta$. Since all our theoretical values are averaged over the angle $\varphi$, we will only consider the out-of-plane tilt. This is done by rotating the mass profile by the angle \(\beta\) around the \(y\)-axis. 
The resulting mass profile is:
\begin{equation}
\begin{split}
\rho\indice{G 2}(r, \theta, \varphi)&=\rho\indice{0,G 2} \exp \left\{-r\left\{\left[\frac{\sin \theta \cos \varphi \cos \beta\indice{G 2}-\cos \theta \sin \beta\indice{G 2}}{x\indice{0,G 2}}\right)^{2}+\left(\frac{\sin \theta \sin \varphi}{y\indice{0,G 2}}\right)^{2}\right]^{2}\right.\\ &+\left.\left.\left(\frac{\sin \theta \cos \varphi \sin \beta\indice{G 2}+\cos \theta \cos \beta\indice{G 2}}{z\indice{0,G 2}}\right)^{4} \right\}^{1 / 4} \right\} .
\end{split}
\end{equation}
The characteristic parameters have been measured by infrared \cite{Dwek1995} and microlensing observations \cite{Novati2008} and are shown in Table~\ref{tab:parameter_bulge}. 

\begin{table}[t]
\caption{This table shows the fixed parameters which are used in the baryonic models for the bulge. The parameters are taken from microlensing \cite{Novati2008} and infrared observations \cite{Dwek1995}}
\label{tab:parameter_bulge}
\centering

\begin{tabular}{c|ccccc}
 & \(\rho_0\left[\mathrm{kg/pc^3}\right]\) & $x_0\left[\mathrm{pc}\right]$ & $y_0\left[\mathrm{pc}\right]$ & $ z_0\left[\mathrm{pc}\right]$ & $\beta \left[\mathrm{degree}\right]$ \\
\hline
E2 & $1.909410\ten{31}$ & $740$ & $160$ & $270$ & $0.6$\\
G2 & $4.7736\ten{30}$ & 1580 & 620 & 430 & 0.7\\
\hline
\end{tabular}
\end{table}
In the second model, denoted by E2, the bulge has an exponential dependency along all axes \cite{Novati2008}:
\begin{align}
\rho\indice{E2}(x, y, z)=\rho\indice{0,E2} \exp \left(-r\indice{E2}\right),
\end{align} 
with
\begin{align}
r\indice{E2}=\left[\left(\frac{x}{x\indice{0,E2}}\right)^{2}+\left(\frac{y}{y\indice{0,E2}}\right)^{2}+\left(\frac{z}{z\indice{0,E2}}\right)^{2}\right]^{1 / 2}.
\end{align}
Considering again the tilt of the bulge, denoted by \(\beta\indice{E2}\), the mass distribution in spherical coordinates is:
\begin{equation}
\begin{split}
\rho\indice{E2}(r,\theta,\varphi)&=\rho\indice{0,E2} \exp \left\{-r\left[\left(\frac{\sin \theta \cos \varphi \cos \beta\indice{E2}-\cos \theta \sin \beta\indice{E2}}{x\indice{0,E2}}\right)^{2}+\right.\right.
 \\ &+ \left.\left.\left(\frac{\sin \theta \sin \varphi}{y\indice{0,E2}}\right)^{2}+\left(\frac{\sin \theta \cos \varphi \sin \beta\indice{E2}+\cos \theta \cos \beta\indice{E2}}{z\indice{0,E2}}\right)^{2} \right]^{1 / 2} \right\}.
\end{split}
\end{equation}
For the characteristic parameters we take values shown in Table~\ref{tab:parameter_bulge}, which are taken from mircolensing \cite{Novati2008} and infrared observations \cite{Dwek1995}. 

\subsection{Disk}
 
The second component is the disk. As for the bulge, we take into account two different models, one modelling the galactic disk as a single disk (denoted as SM) and the other one modelling it as the superposition of a thin and a thick disk (TT). 

The SM model can be described by the following mass distribution \cite{McMillan2018}
\begin{equation}
\rho\indice{D}(R, z)=\frac{\Sigma\indice{0,D}}{2 z\indice{D}} \exp \left(-\frac{|z|}{z\indice{D}}-\frac{R}{R\indice{D}}\right).
\label{eq:disk}
\end{equation}
where \(z\indice{D}\) is a scale height for the thickness, \(R\indice{D}\) is a scale length for the radius of the disk and \(\Sigma\indice{0,D}\) is the central surface density.

\begin{table}[t]
\caption{This table shows the fixed parameters which are used in the baryonic models for the disk.}
\label{tab:parameters_disk}
\centering

\begin{tabular}{c|ccccccccc}
 & \(\Sigma_0\left[\mathrm{kg/pc^2}\right]\) & $z\left[\mathrm{pc}\right]$ & $R\left[\mathrm{kpc}\right]$ & $ \rho_0\left[\mathrm{kg/pc^3}\right]$ & $h_1 \left[\mathrm{pc}\right]$ & $h_2 \left[\mathrm{pc}\right]$ & $H \left[\mathrm{kpc}\right]$ & $\beta $ & $R_\odot \left[\mathrm{kpc}\right]$\\
\hline
SM & $1.0144\ten{32}$ & 400 & 2.15 &  & \\
TT &  &  &  & $0.0981\ten{30}$ & 270 & 440 & 2.75 & 0.565 & 8\\
\hline
\end{tabular}
\end{table}
For the TT model, we use the superposition of two disks, which in spherical coordinates reads \cite{CheonghoHan}
\begin{equation}
\begin{split}
\rho\indice{TT}(r, \theta)&=\rho\indice{0,TT} \frac{1}{\eta(r, \theta)} \exp \left(-\frac{r \sin \theta-R_{\odot}}{H\indice{TT}}\right)\\
&\cdot\left[\left(1-\beta\indice{TT}\right) \mathrm{sech}^{2} \left(\frac{r \cos \theta}{\eta(r, \theta) h\indice{1,TT}}\right)+\beta\indice{TT} \exp \left(-\frac{r|\cos \theta|}{\eta(r, \theta) h\indice{2,TT}}\right)\right],
\end{split}
\end{equation}
with
\begin{equation}
\eta(r, \theta)=\max \left\{\frac{r \sin \theta}{9025 \einheit{p c}}+0.114,0.670\right\},
\end{equation}
and \(H\indice{TT}\) being the scale length for the radius, \(h\indice{1 , TT},h\indice{2 , TT}\) being the scale heights for the thickness and \(R_\odot\) being the radius from the galactic center to the sun.

The characteristic parameters of the two models, shown in Table~\ref{tab:parameters_disk}, are chosen following \cite{J.Bovy2013} and \cite{CheonghoHan}.

\subsection{Gas}

The gas component, distributed throughout the whole galaxy, can be separated into three regions.
A detailed description of the three regions can be found in \cite{Ferriere2012,Ferriere2007,Ferriere1998}.

The first region describes the gas up to \(10\einheit{pc}\). Here the gas can be modelled as a point mass, such that the gravitational potential is \cite{Ferriere2012}:
\begin{equation}
\Phi\indice{Gas <10\einheit{pc}}=-\frac{G M\indice{Gas < 10\einheit{pc}}}{r}.
\end{equation}
The mass is estimated to be around \(M\indice{Gas <10\mathrm{pc}}=7\times 10^5\,\mathrm{M_\odot}\)

For the second region, which describes all the gas inside \(2\einheit{kpc}\), we assume that the gas mostly consist of hydrogen gas in molecular (H 2) atomic (H 1) or ionised (H +) form. Therefore, the number density of hydrogen is \cite{Ferriere2007}:
\begin{equation}
\langle n\indice{H}\rangle=2\langle n\indice{H,2}\rangle+\langle n\indice{H,1}\rangle+\langle n\indice{H^+}\rangle .
\label{eq:numberdensityH}
\end{equation}
By using the ratios of the mass of other gas components relative to the hydrogen mass, we can relate the mass density of the total gas to the number density of the hydrogen \cite{Ferriere2007}.
\begin{equation}
\rho\indice{Gas <2\einheit{kpc}}= 1.453 \rho\indice{H}=1.453 m\indice{p}\langle n\indice{H}\rangle ,
\end{equation}
In this case, we neglected the mass of the electron and replaced the hydrogen mass with the mass of a proton (\(m\indice{p}\)). The structure of the number density can be found in \cite{Ferriere2007}.

The last region describes all the gas outside of \(2\einheit{kpc}\). The gas exist in molecular (m), cold neutral (c), warm neutral (w), warm ionized(i) and hot ionized (h) form \cite{Ferriere1998}.
\begin{equation}
\langle n\indice{H} \rangle = \langle n\indice{m} \rangle +\langle n\indice{c} \rangle +\langle n\indice{w} \rangle +\langle n\indice{i} \rangle +\langle n\indice{h} \rangle
\label{eq:hydrogenabove2}
\end{equation}
In this case, only Hydrogen and Helium are considered. The latter can be again related to Hydrogen, such that \cite{Ferriere1998}
\begin{equation}
\rho\indice{Gas  > 2\einheit{kpc}}=1.36 \langle n\indice{H} \rangle m\indice{p},
\label{eq:numberdensitygasabove2}
\end{equation}
As in the previous cases, the mass of the electron can be neglected and the number density of the hydrogen can be found in \cite{Ferriere1998}.

\begin{figure}[t]
\centering
\includegraphics[scale=0.5]{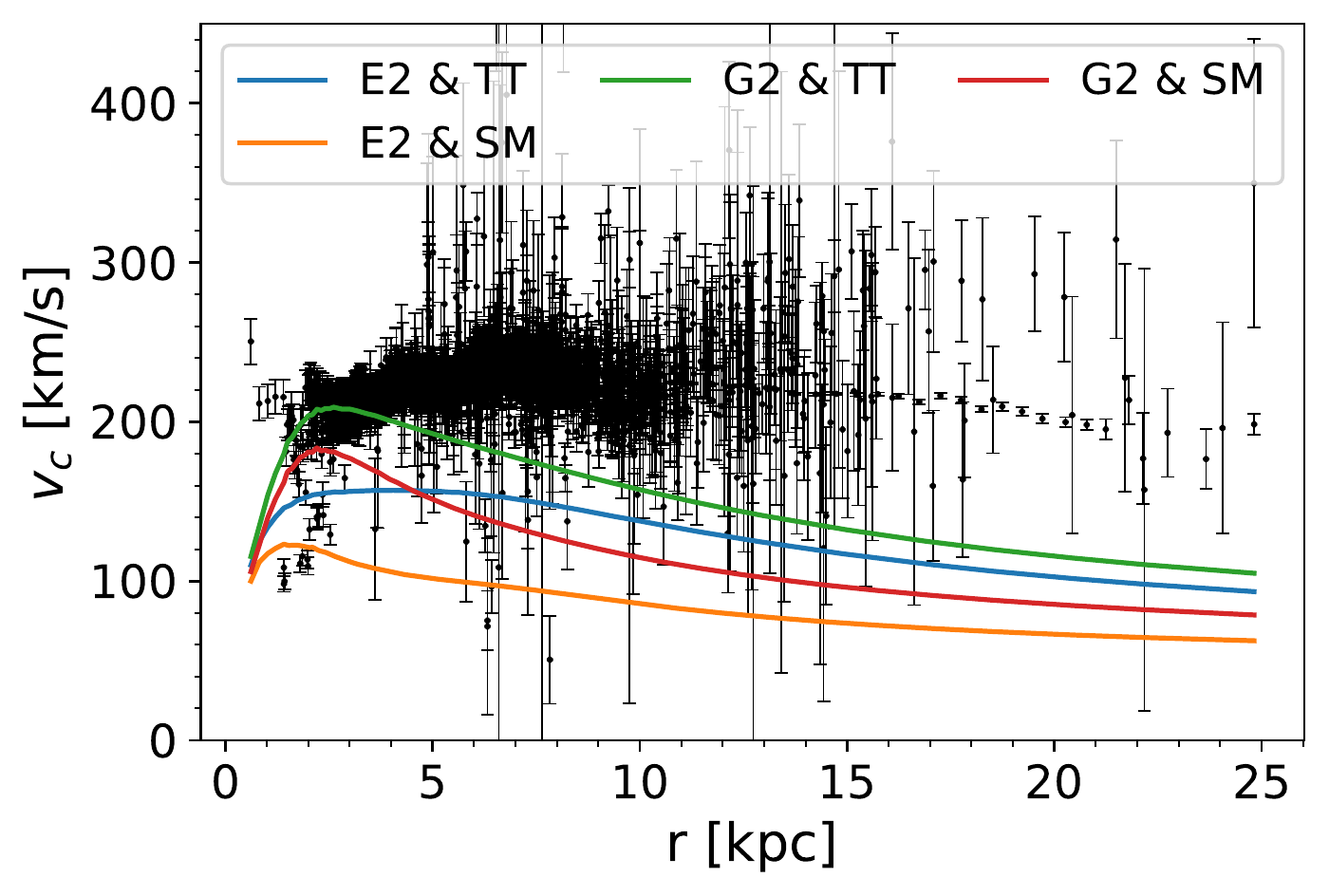}
\caption{Rotation curves for only baryonic matter of a source of gravity. The baryonic matter alone is not sufficient to support the observed velocities beyond $\sim$5kpc, \cite{Iocco:2015xga}.}
\label{fig:baryonic_models}
\end{figure}
\begin{figure}[t]
\centering
\includegraphics[scale=0.5]{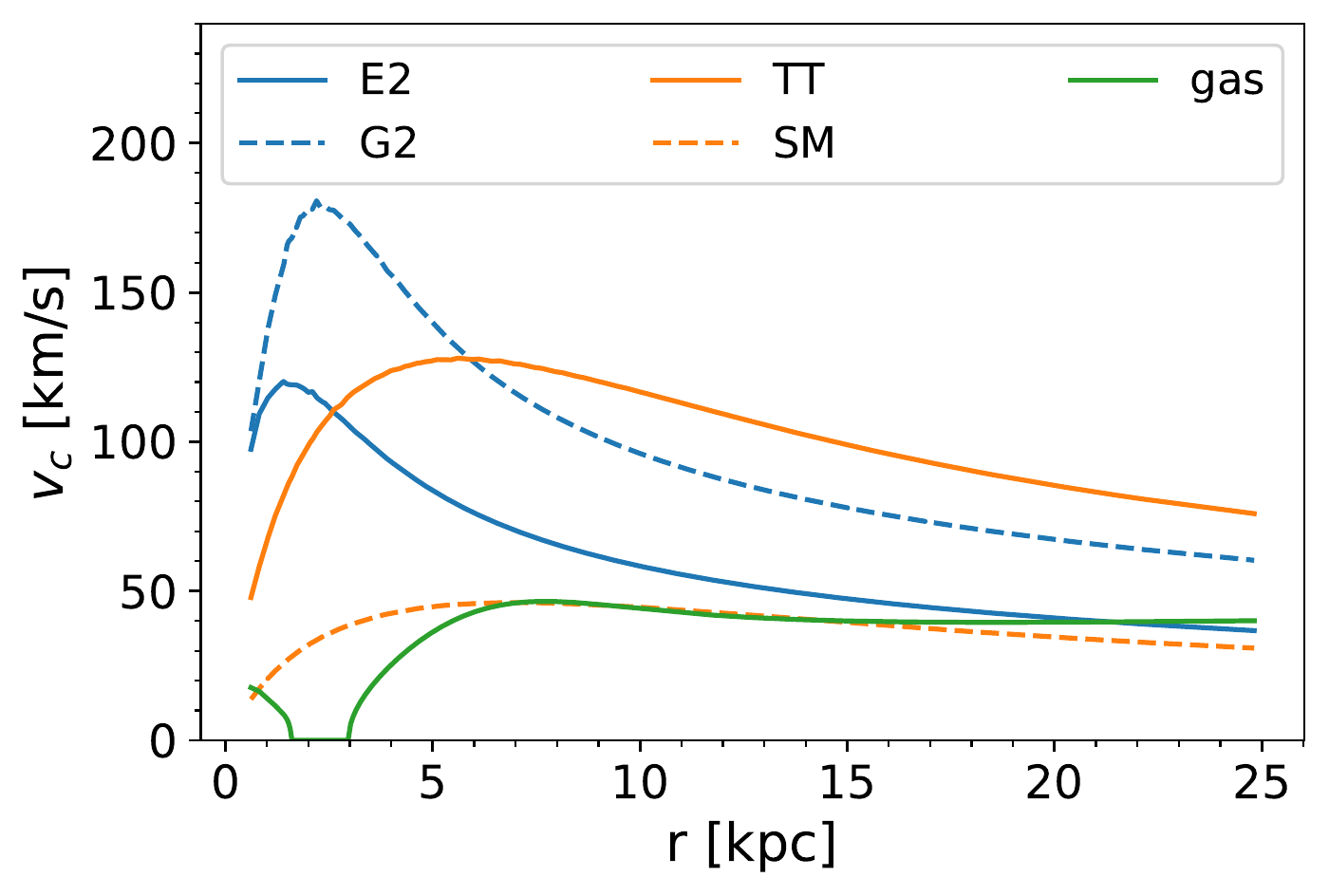}
\caption{Rotation curves due to the different components of the baryonic parts of the Milky Way.}
\label{fig:baryonic_components}
\end{figure}

\vspace{0.65cm}

The combination of all the described models leads to four baryonic morphologies, with no free parameters.
It is worth mentioning that all morphologies are derived from observations of stars in the Galactic disk and bulge, and are therefore purely empirical.  Hence, there is no ``a priori'' modelization (the term model in the context of baryons has been used only for simplicity) of the baryonic components, whose corresponding rotation curves are shown in Fig.~\ref{fig:baryonic_models}. We notice two main features: {\it a)} none of the morphologies considered is able to reproduce the observed rotation curve, unless  an additional component is included already within the solar circle \cite{Iocco:2015xga}; {\it b)} the rotation curves from the different morphologies differ substantially: this is to be imputed to the different normalization and shapes of stellar bulge and disk as inferred from different observational groups. This uncertainty still represents a major one in the computation of the Galaxy's gravitational potential \cite{Pato:2015dua,Benito:2019ngh,Benito:2020lgu}.For what concern our study, our conclusions stay unchanged whichever baryonic morphology is adopted.

%In this plot, it can be clearly noticed that the only baryonic matter with the measured parameters is not sufficient to describe the rotation curve of the milky way galaxy. How each baryonic component influences the rotation curve is shown in figure \ref{fig:baryonic_components}.

\subsection{Eilers et al. models\label{subs:eilers}}

In addition to the previous baryonic morphologies, we consider an extra one, employed in \cite{Eilers2018} together with the NFW profile to fit the rotation curve to their  circular velocity data. This model describes the baryonic part of the galaxy by including only the bulge and the disk. For the description of the bulge a Plummer potential is adopted \cite{2017A&A...598A..66P}:
\begin{align}
\Phi\indice{bulge}(r)=-\frac{G M\indice{bulge}}{\left(r^{2}+b\indice{bulge}^{2}\right)^{1 / 2}}    
\end{align}
The galactic disk is modelled as a superposition of a thin and a thick disk. Both disks are described with the same gravitational potential \cite{2017A&A...598A..66P}:
\begin{align}
\Phi\indice{disk}(R, z)=-\frac{G M\indice{disk}}{\left(R^{2}+\left[a\indice{disk}+\sqrt{z^{2}+b\indice{disk}^{2}}\right]^{2}\right)^{1 / 2}}
\end{align}
The matching parameters for the potentials of the different components are taken from \cite{2017A&A...598A..66P} (Model 1) and are shown in Table~\ref{tab:eilers_model}.

\begin{table}[t]
\caption{This table shows the fixed parameters which are used for the baryonic model of Eilers et al.. The values are taken from \cite{2017A&A...598A..66P} (Model 1).}
\label{tab:eilers_model}
\centering

\begin{tabular}{c|ccc}
 & M $\left(2.32\ten{7} M_\odot\right)$ & a [kpc] & b [kpc]\\
\hline
bulge & 460.0 &  & 0.3000 \\
thin disk & 1700.0 & 5.3000 & 0.25 \\
thick disk & 1700.0 & 2.6 & 0.8 \\
\hline
\end{tabular}
\end{table}

\subsection{Dark Matter model}
For the description of the dark matter halo, we used the NFW mass density profile \cite{Coe2010}:
\begin{align}
\rho(r)=\frac{\rho\indice{s}}{\left( \frac{r}{r\indice{s}} \right) \left( 1+\frac{r}{r\indice{s}} \right)^2}.
\label{eq:NFW}
\end{align}
For small radii the density scales with $\rho \propto r^{-1}$. This means that it increases fast and can be described as a cusp. At large radii the density scales with $\rho \propto r^{-3}$ \cite{Coe2010}.

We can also write the parameter \(r_s\) and \(\rho_s\) in terms of the parameters \(c\) and \(M\indice{200}\), where \(c\) is the concentration parameter, which is defined by \(c\equiv r\indice{200}/r\indice{s}\). Defining the radius \(r\indice{200}\) as the radius at which the average overdensity is 200 times the critical density, the parameter \(M\indice{200}\) is defined as the virial mass of the dark matter halo within \(r\indice{200}\). The relations between the parameters are \cite{1996NFW}
\begin{align}
\rho\indice{s}&=\frac{200}{3}\frac{c^3 \rho\indice{crit}}{\log(1+c)-\frac{c}{1+c}}
\label{eq:c}\\
r\indice{s}&=\frac{1}{c}\left(\frac{3M\indice{200}}{4\pi 200 \rho\indice{crit}}\right)^{1/3}.
\label{eq:m200}
\end{align}
For the calculation of the parameters we used the value \(\rho\indice{crit}=143.84\phantom{,}M_\odot\einheit{kpc^{-3}}\) \cite{Almeida2018} for the critical density and the errors are calculated with standard error propagation.

%%%%%%%%%%%%%%%%%%%%%%%%%%%%%%%%%%%%%%%
%%%%%%%%%%%%%%%% Sec IV %%%%%%%%%%%%%%%
%%%%%%%%%%%%%%%%%%%%%%%%%%%%%%%%%%%%%%%
\section{Circular velocity}\label{section:circular_velocity}

The circular velocity can be derived from the potential as follows,
\begin{equation}
v^2\indice{c}=r \totalderiv{\Psi}{r}\,.
\label{eq:vc}
\end{equation}
Due to the linearity of the derivative, it is possible to split the right hand side into the velocities due to each component of the the galaxy:
\begin{equation}
v^2\indice{c}=v^2\indice{Bulge}+v^2\indice{Gas}+v^2\indice{Disk}+v^2\indice{DM}\,.
\label{eq:totalvc}
\end{equation}
Moreover, we assume that only dark matter is effectively coupled to modified gravity (that is, we neglect any possible small coupling of baryons), such that only \(v^2\indice{DM}\) has to be calculated with the modified gravitational potential (see Eq.~\eqref{eq:modgravpot}).
\begin{equation}
v^2\indice{DM}=v^2\indice{N}+v^2\indice{Y}=r \totalderiv{\Psi\indice{N}}{r}+r \totalderiv{\Psi\indice{Y}}{r}
\end{equation}
The index \(N\) describes the Newtonian and the index \(Y\) describes the Yukawa part of the gravitational potential.

The gravitational potentials for the baryonic parts of the galaxy are calculated numerically with a multipole expansion \cite{J.Binney2008}. This has been done for several different values of $r$. 
The resulting values are then interpolated to get the gravitational potential as a function of $r$ for each baryonic model. 
These functions of $r$ allow us to calculate the the corresponding velocities which are then used in the fitting process.

To solve Eq.~\eqref{eq:modgravpot} for the dark matter part, we split up the integral into the Newtonian and Yukawa part and we solve them separately. The Yukawa part can be written as \cite{Amendola2019}
\begin{align}
\Psi\indice{Y}(r)&=-2\pi G\int^\infty_0 d r^\prime \phantom{.}r^{\prime^2} \rho(r^\prime) F(r^\prime,r)
\end{align}
\begin{equation}
F(r^\prime,r)=
\begin{cases} 
      \frac{\e^{-m(r^\prime-r)}-\e^{-m(r^\prime+r)}}{m r^\prime r} & r^\prime>r \\
      \frac{\e^{m(r^\prime-r)}-\e^{-m(r^\prime+r)}}{m r^\prime r} & r^\prime<r 
\end{cases}
\end{equation}
The integral can be obtained analytically. The solution is \cite{Almeida2018}:
\begin{equation}
\begin{split}
\Psi_{\mathrm{Y},\mathrm{NFW}}=&-\frac{2\pi\mathrm{G}\beta\rho\indice{s} r^3\indice{s}}{r}   \left[   \e^{m(r\indice{s}+r)}   \Ei{-m(r\indice{s}+r)}   -   \e^{-m(r\indice{s}+r)}   \Ei{m r\indice{s}}\right. \\
&- \left. \e^{m(r\indice{s}-r)}   \Ei{-m r\indice{s}}   +   \e^{-m(r\indice{s}+r)}   \Ei{m(r\indice{s}+r)}   \right]
\end{split}
\label{eq:yukawaNFW}
\end{equation}
with the exponential integral function:
\begin{align}
\Ei{x}=-\int^{\infty}_{-x}\frac{\e^{-t}}{t}\mathrm{d}t
\end{align}
The same can be done for the Newtonian part of the potential, obtaining
\begin{align}
\Psi_{\mathrm{N},\mathrm{NFW}}=-\frac{4\pi\mathrm{G}\rho\indice{s} r^3\indice{s}}{r}\log{\frac{r\indice{s}+r}{r\indice{s}}}.
\label{eq:newtNFW}
\end{align}
Hence, the circular velocity due to dark matter can be calculated with Eq.~\eqref{eq:vc}. We obtain for the Newtonian part
\begin{equation}
v^2\indice{N,NFW}=\frac{4\pi\mathrm{G}\rho\indice{s}r^3\indice{s}}{r}\left(\log\left[1+\frac{r}{r\indice{s}}\right]-\frac{r}{r+r\indice{s}}\right)\,,
\label{eq:vcNNFW}
\end{equation}
and for the Yukawa part
\begin{equation}
\begin{split}
v^2\indice{Y,NFW}= &-\frac{2\pi\mathrm{G}\beta\rho\indice{s}r^3\indice{s}}{r}\left[\frac{2 r}{r\indice{s}+r}+\e^{\frac{r\indice{s}+r}{\lambda}}\left(\frac{r}{\lambda}-1\right)\Ei{-\frac{r\indice{s}+r}{\lambda}}\right.\\
&+ \left.\e^{-\frac{r\indice{s}+r}{\lambda}}\left(1+\frac{r}{\lambda}\right)\left(\e^{\frac{2r\indice{s}}{\lambda}}\Ei{-\frac{r\indice{s}}{\lambda}}+\Ei{\frac{r\indice{s}}{\lambda}}-\Ei{\frac{r\indice{s}+r}{\lambda}}\right) \right] \,,
\end{split}
\end{equation}
respectively.

%%%%%%%%%%%%%%%%%%%%%%%%%%%%%%%%%%%%%%%
%%%%%%%%%%%%%%%% Sec V %%%%%%%%%%%%%%%%
%%%%%%%%%%%%%%%%%%%%%%%%%%%%%%%%%%%%%%%
\section{Circular velocity data}\label{section:data}

We analysed data from several different groups. Very precise measurements have been published recently in \cite{Eilers2018}; this dataset will be referred to as Eilers et al.. They used data from Gaia, WISE, 2MASS and APOGEE. In total 23129 red giant stars inside the disk of the Milky Way were measured to calculate the rotation curve. The other larger dataset (to be denoted {\it galkin}) is a collection \cite{Iocco:2015xga,MiguelPato2016} of several measurements of the kinematics of gas, stars and masers inside our Galaxy. Due to the radial uncertainty, the velocity error for the {\it galkin} dataset appears underestimated. To compensate for this, in our analysis  we take a minimum threshold error of 5\% for all circular velocities from the {\it galkin} dataset. 

\begin{figure}[t]
\begin{subfigure}{.5\textwidth} 
\centering
\includegraphics[scale=0.6]{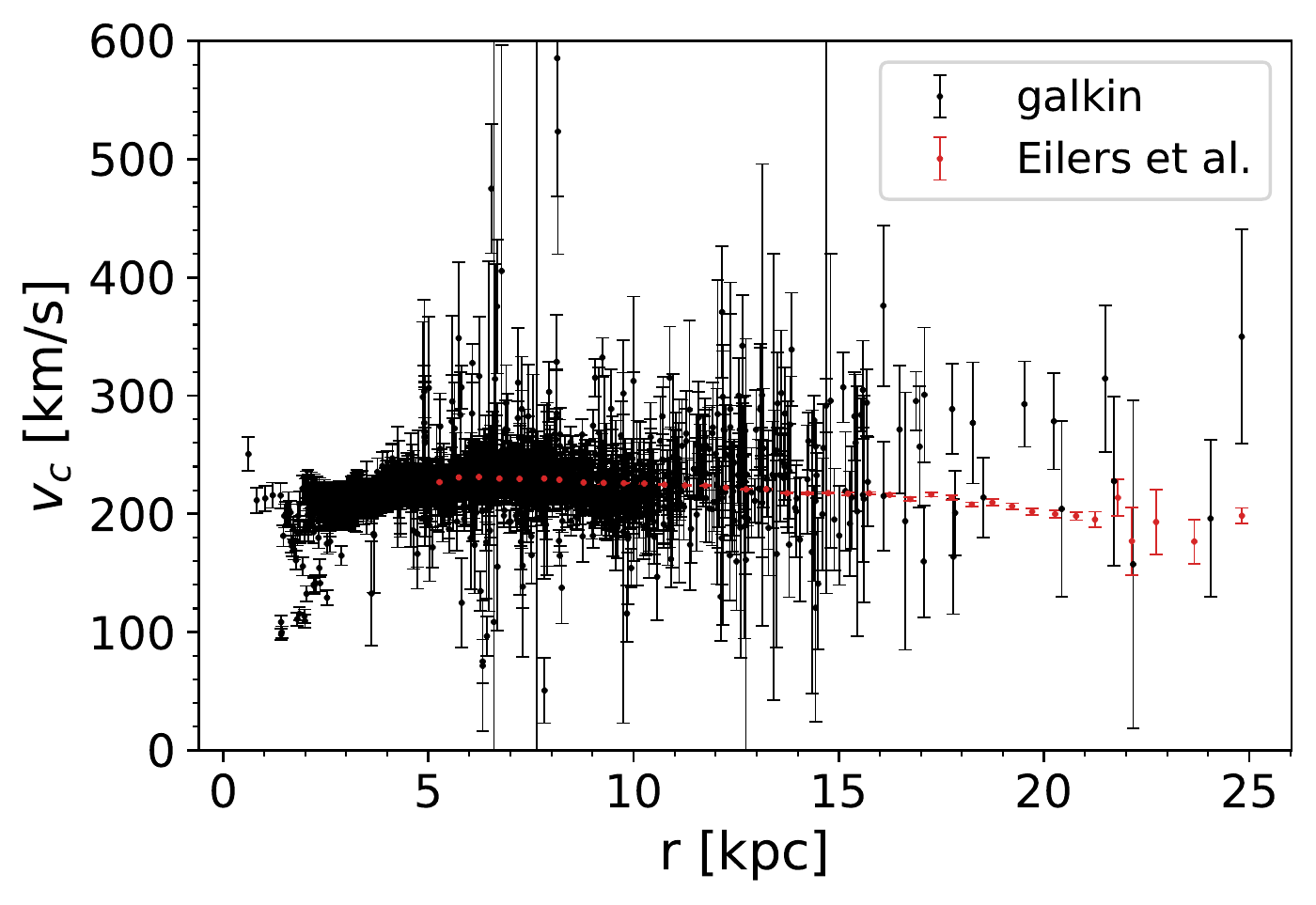} 
\end{subfigure} 
\begin{subfigure}{.5\textwidth} 
\centering
\includegraphics[scale=0.6]{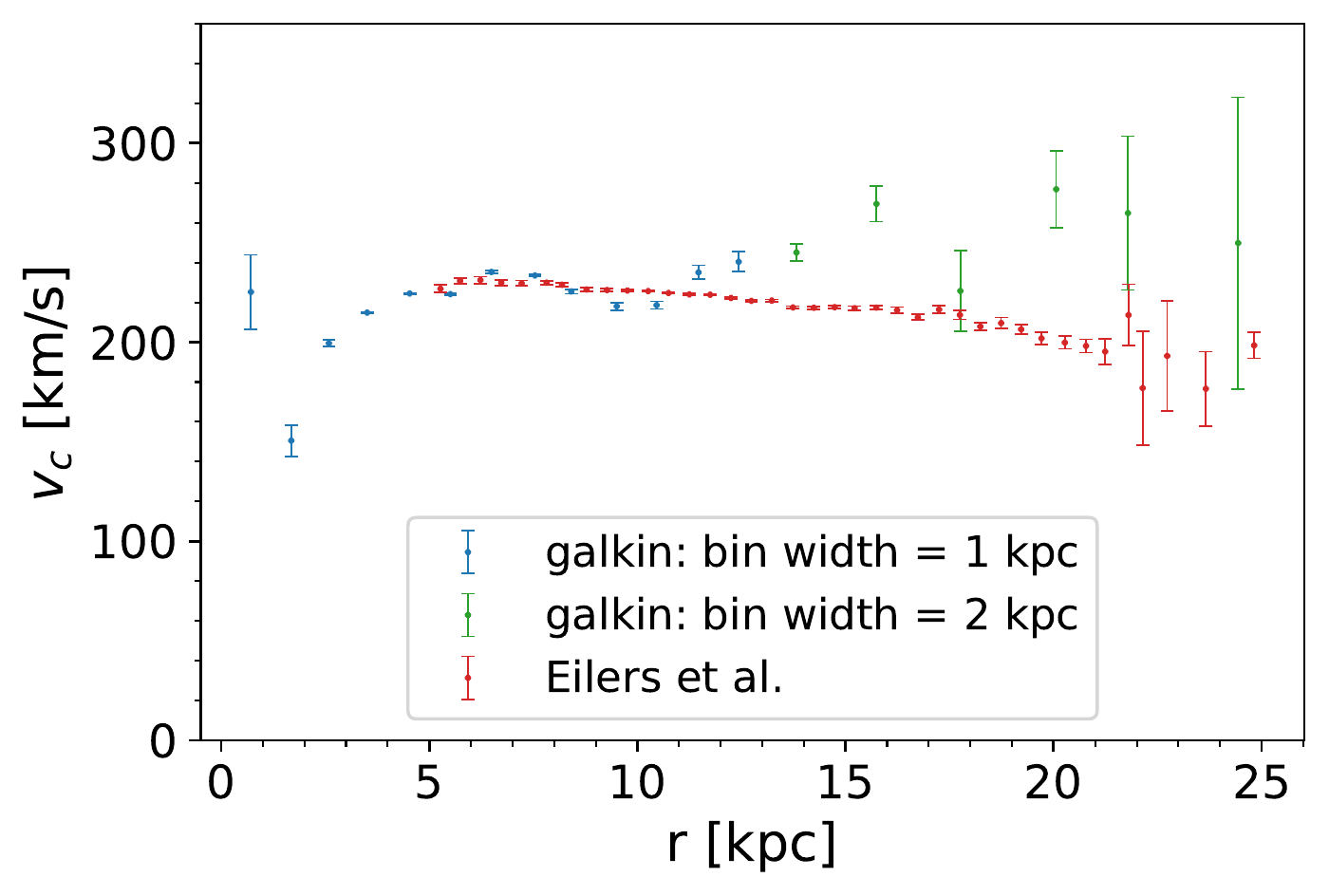} 
\end{subfigure} 
\caption{ The two plots are showing the two different data sets used for this analysis. For the bottom plot, the data from {\it galkin} is binned.}
\label{fig:data}
\end{figure}

The two datasets are shown in Fig.~\ref{fig:data}. In the numerical analysis, the data are left unbinned but, for a simpler representation, we also show the binned data in the bottom panel of Fig.~\ref{fig:data}.  The resulting data points shown in Fig.~\ref{fig:data} are the mean of the radius and the weighted mean of the circular velocity. The errorbars represent the standard deviation of the binned data points\footnote{The different datasets have been proven to be compatible, \cite{Karukes:2019jxv} and the choice of different datasets (even when adopting historical compilations, updated with modern data \cite{Sofue:2020rnl})
does not alter the conclusions of studies like the ones performed here, even when compared to newest datasets like those provided by Gaia.}.

It is to be noticed that the so-called ``Rotation Curve method'' described above, intrinsically relies on a series of assumptions, namely that:
{ \it(a)} the gravitational potential has a cylindrical symmetry;
{\it(b)} the adopted tracers are on orbits that follow the gravitational potential;
{\it(c)} the orbits of the adopted tracers are circular. 
Whereas all of these assumptions are known to be true to a good degree in regions of the Galaxy above 3-4 kpc, condition {\it a)} is well known to be broken within the central 3kpc, because of the departure from cylindrical symmetry of the Galactic Bulge, and it is still highly debated whether {\it b)} and {\it c)} are true for the objects that may be adopted as tracers within that region.
For these reasons, most analysis tend to discard {\it a priori} the data within the central 3kpc in Rotation Curve analysis.
In fact, some analyses of the Galactic Center prefer to rely on a complete Jeans analysis of the potential, based on the complex motion of stars within that region, see e.g. \cite{Hooper:2016ggc,Iocco:2016itg} and references therein.

 {\it A posteriori}, one can also notice that the data in Fig.~\ref{fig:data} show an intrinsic tension at $r\approx 1-3$ kpc, pointing to both an high ($\approx 200$ km/sec) and a low ($\approx 100-150$ km/sec) velocity.
These data have been mostly measured by two groups, \cite{BurtonGrodon78} and \cite{Knapp85}, using the same technique, CO terminal velocities.
As we will discuss later, this feature is suspicious and the results should be carefully processed.

%%%%%%%%%%%%%%%%%%%%%%%%%%%%%%%%%%%%%%%
%%%%%%%%%%%%%%%% Sec VI %%%%%%%%%%%%%%%
%%%%%%%%%%%%%%%%%%%%%%%%%%%%%%%%%%%%%%%
\section{Maximum likelihood fit and confidence regions}\label{section:maximum_likelihood}

For the fit, we adopt the maximum posterior method with flat priors. The posterior coincides with the likelihood inside the parameter range. The likelihood has the following form:
\begin{equation}\label{eq:lik}
L(\rho\indice{s},r\indice{s},\beta, \lambda)=\exp\left[-\frac{1}{2}\sum_i \frac{(v^2_{\mathrm{c}, i}-v^2_{\mathrm{b},i}-v^2_{\mathrm{dm},i}-v^2_{\mathrm{mod},i})^2}{4v_{\mathrm{c}, i}^2\sigma^2_{\mathrm{v}, i}} \right]\,,
\end{equation}
where we assume that the errors of the rotation curve follow a Gaussian distribution and we use as data the circular velocity squared. The subscript ``c" describes the data, ``b" is the baryonic component, 
while ``dm" and ``mod" are the Newtonian and Yukawa contribution to the circular velocity, respectively. 
The denominator of Eq.~\eqref{eq:lik} is the squared error of the squared velocity, \(\sigma\indice{v^2}=2v\indice{c}\sigma\indice{v}\).

The free parameters are then ${\beta,\lambda,r_s,\rho_s}$. For the priors we set the following ranges:
\begin{align*}
-15 <\,\, &\beta < 15\,; \\
0.1\,\mathrm{kpc} <\,\, &\lambda < 30\,\mathrm{kpc} \,;\\
0.1\,\mathrm{kpc} <\,\, &r\indice{s} < 50\,\mathrm{kpc} \,;\\
\ten{35}\,\mathrm{kg/kpc^3} <\,\, &\rho\indice{s} < 10^{40}\,\mathrm{kg/kpc^3} \,.
\end{align*}

For the coupling $\beta$ we include the possibility of both a repulsive or an attractive Yukawa correction.  For the Yukawa scaling length $\lambda$ we take a maximal value, 
larger that the whole range of our dataset.
For the two characteristic parameters of the NFW profile 
 we consider a broad interval around values found by previous analysis \cite{Eilers2018, Lin2019}. 
A minimal finite value for \(r\indice{s}\) and $\lambda$ is employed to avoid divergences.

We adopt a MonteCarlo random sampling method to scan the parameter space. To create the MonteCarlo Markov chain (MCMC),  we employ the python package {\tt emcee} \cite{Foreman-Mackey2013}; while the python package {\tt getdist} \cite{2019arXiv191013970L} is used to create the marginalised plots and to calculate the errors. To estimate the errors of the fit parameters, we take the \(1 \sigma\) confidence region. 
For the standard gravity case we consider $\beta =0$ as fixed parameter, while the two remaining parameters $r\indice{s}$ and $\rho\indice{s}$ are fitted as described above. As a prior we used a uniform prior for the linear parameter range. We also tried a logarithmic prior, but found no differences to the linear prior. The modified gravity case is fitted by also using $\beta$ and $\lambda$ as free fit parameters. To get a better result for parameters with a possible wide range in the parameter space (all except \(\beta\)), the logarithm is now used in the fitting process. This means that we use for all parameters except $\beta$ a logarithmic parameter range with a uniform prior. For $\beta$ we use a linear parameter range with a uniform prior. Otherwise, the fitting procedure is the same as for the standard gravity case.

The convergence of the MCMC chain is a key issue. To monitor this, we use the auto-correlation time $(\tau)$ of the chain. To get trustworthy estimates  
of it, we use chains with at least \(N=50\cdot \tau\) steps, as suggested by the developers \cite{2020}. 
As general rule to ensure a good convergence of the chain, we use more than \(N=50\cdot\tau\) steps per run. 
Moreover, we remove always 10\% of our samples as a burn-in: the first steps of the chain are used to find the peak, so we can discard them.

Another parameter to take care of is the acceptance fraction.  
Following the developers' suggestion \cite{Foreman-Mackey2013}, we ensure the acceptance fraction for all chains to be between 0.2 and 0.5.

The {\tt emcee} sampler gives the possibility to run multiple ``walkers" in one go. These walkers are  different chains, but the proposal distribution for the next step depends on the position of all the other walkers \cite{2020}. So to increase the number of samples per analysis, we used 20 walkers for the modified gravity and 32 for the standard gravity case. For the more complex model (modified gravity) less walkers are used to decrease the runtime of the fit.

%%%%%%%%%%%%%%%%%%%%%%%%%%%%%%%%%%%%%%%
%%%%%%%%%%%%%%% Sec VII %%%%%%%%%%%%%%%
%%%%%%%%%%%%%%%%%%%%%%%%%%%%%%%%%%%%%%%
\section{Model selection}\label{section:model_selection}

To distinguish between the different baryonic models and the cases with and without modified gravity, we perform a model selection test. We use three statistical indicators:    the simple frequentist $\chi^2$ test, the Bayes ratio, and the Bayes Information criteria (BIC). In the following we will compare them in turn.

For $\chi^2$, we use the relation between the maximum likelihood and the minimal $\chi^2$, which is $\chi^2\indice{min}=-2\log\mathcal{L}\indice{max}$. The number of degrees of freedom $f$ is calculated by the difference between the number of data points $N$ and the number of free parameters $k$, such that  $\chi^2\indice{red}=\chi^2\indice{min}/f$.

The Bayes ratio is the ratio of the Bayesian evidences of two models to be compared. We estimate the evidence by approximating the likelihood and the priors as Gaussian. In this case the evidence can be calculated as (see e.g. \cite{Trotta_2008}):
\begin{align}
E=\mathcal{L}_{\max } \sqrt{\frac{\det \mathbf{P}}{\det \mathbf{Q}}} \exp \left[-\frac{1}{2}\left(\hat{\theta}_{\alpha} F_{\alpha \beta} \hat{\theta}_{\beta}+\bar{\theta}_{\alpha} P_{\alpha \beta} \bar{\theta}_{\beta}-\tilde{\theta}_{\alpha} Q_{\alpha \beta} \tilde{\theta}_{\beta}\right)\right],
\label{eq:evidence}
\end{align}
where \(\theta_\alpha=\left\lbrace r\indice{s},\rho\indice{s},\beta,\lambda \right\rbrace\) are the fit parameters. The hat describes the best-fitting parameters, the bar describes the mean of the priors and \(\tilde{\theta}_\alpha=\left(\mathbf{Q}^{-1}\right)_{\alpha \beta}\left[F_{\beta \sigma} \hat{\theta}_{\sigma}+P_{\beta \sigma} \bar{\theta}_{\sigma}\right]\). 
The matrix \(\mathbf{F}\) is the Fisher matrix, \(\mathbf{P}\) is the inverse covariance matrix of the priors, and \(\mathbf{Q}=\mathbf{F}+\mathbf{P}\). Additionally, we assume that the priors are uncorrelated. For the Gaussian variances of the priors, we take the variance of uniform distributions for each parameter
\begin{align}
\sigma^2=\frac{1}{12}\left(b-a\right)^2,
\end{align}
with \(I=[a,b]\) being the interval in which the distribution is non-zero. 
The evidence, Eq.~\eqref{eq:evidence}, can be simplified in the limit of weak priors, which is indeed our case, obtaining
\begin{align}
B_{12}=\frac{E_1}{E_2}= e^{-\frac{1}{2}\left(\chi_{\min , 1}^{2}-\chi_{\min , 2}^{2}\right)} \sqrt{\frac{\operatorname{det} \mathbf{P}_{1} \operatorname{det} \mathbf{F}_{2}}{\operatorname{det} \mathbf{P}_{2} \operatorname{det} \mathbf{F}_{1}}} \,.
\end{align}
Moreover, comparing the baryonic models, the matrix $\mathbf{P}$ for both models is identical, hence the equation for the Bayes ratio simplifies further as
\begin{align}\label{eq:B12}
B_{12}=e^{-\frac{1}{2}\left(\chi_{\min , 1}^{2}-\chi_{\min , 2}^{2}\right)} \sqrt{\frac{\det \mathbf{F}_{2}}{\det \mathbf{F}_{1}}} \,.
\end{align}
If one further takes $\mathbf{F}_{1}\approx \mathbf{F}_{2}$, then $-2\log B_{12}$ reduces to the $\chi^2$ estimator.

Looking at Eq.~\eqref{eq:B12}, if $B_{12}$ is larger than one,  Model 1 is favoured by the data, otherwise Model 2 is favoured. A popular way to express the preference in a more familiar language
 is provided by the Kass-Raftery scale \cite{Kass1995}.
For instance, according to this scale, if $2\log\indice{e}B_{12}$ is larger than 6, Model 1 is ``strongly" favoured.

Alternatively, using the standard Gaussian language, the Bayesian probability of Model 1 being  the correct model is described by 
\begin{equation}
P=\frac{B_{12}}{1+B_{12}}\,.
\end{equation}
For example, if $P>0.95$, i.e. $2\log B_{12}>5.88$, then Model 1 should be preferred at 2$\sigma$ over Model 2.

Lastly, the BIC can be simply calculated with the following formula \cite{schwarz1978}:
\begin{align}
\mathrm{BIC}=-2 \ln \mathcal{L}_{\max }+ k \ln N\,,
\end{align}
where $k$ is the number of fit parameters and $N$ is the number of data points. The model with the smallest BIC value is the best-fitting one. 
Differences between the BIC values will suggest a preference for one model in relation to another.

The BIC can be obtained from the Gaussian Bayes' ratio if one neglects the priors and assumes  the Fisher matrices to be diagonal and scaling as  $\sigma_0 N^{1/2}$, where $\sigma_0$ is a constant. In this case, in fact, $-2\ln (\det \mathbf{F})^{1/2}=k\ln N+const$.
For simplicity, below we make mostly use of the BIC.

%%%%%%%%%%%%%%%%%%%%%%%%%%%%%%%%%%%%%%%
%%%%%%%%%%%%%% Sec VIII %%%%%%%%%%%%%%%
%%%%%%%%%%%%%%%%%%%%%%%%%%%%%%%%%%%%%%%
\section{Results}\label{section:results}

This section is organized into three parts. In the first one, we show and discuss the results from the analysis with a 3 kpc cut-off for the data; the second part contains the results from the analysis with a 5 kpc cut-off for the data; in the third part, for completeness, we use the full data set for the analysis.

\subsection{Cut-off at 3 kpc}
For the analysis in this part, we neglect all data points with a radius smaller than 3 kpc, according to the motivations discussed above. 
We present here both the results obtained assuming standard gravity, i.e. $\beta = 0$,  and then, the ones when we assume modified gravity, i.e. $\beta$ and $\lambda$ as free fit parameters.

\subsubsection{Standard gravity}
\begin{figure}[t]
\begin{subfigure}{.5\textwidth} 
\centering
\includegraphics[scale=0.6]{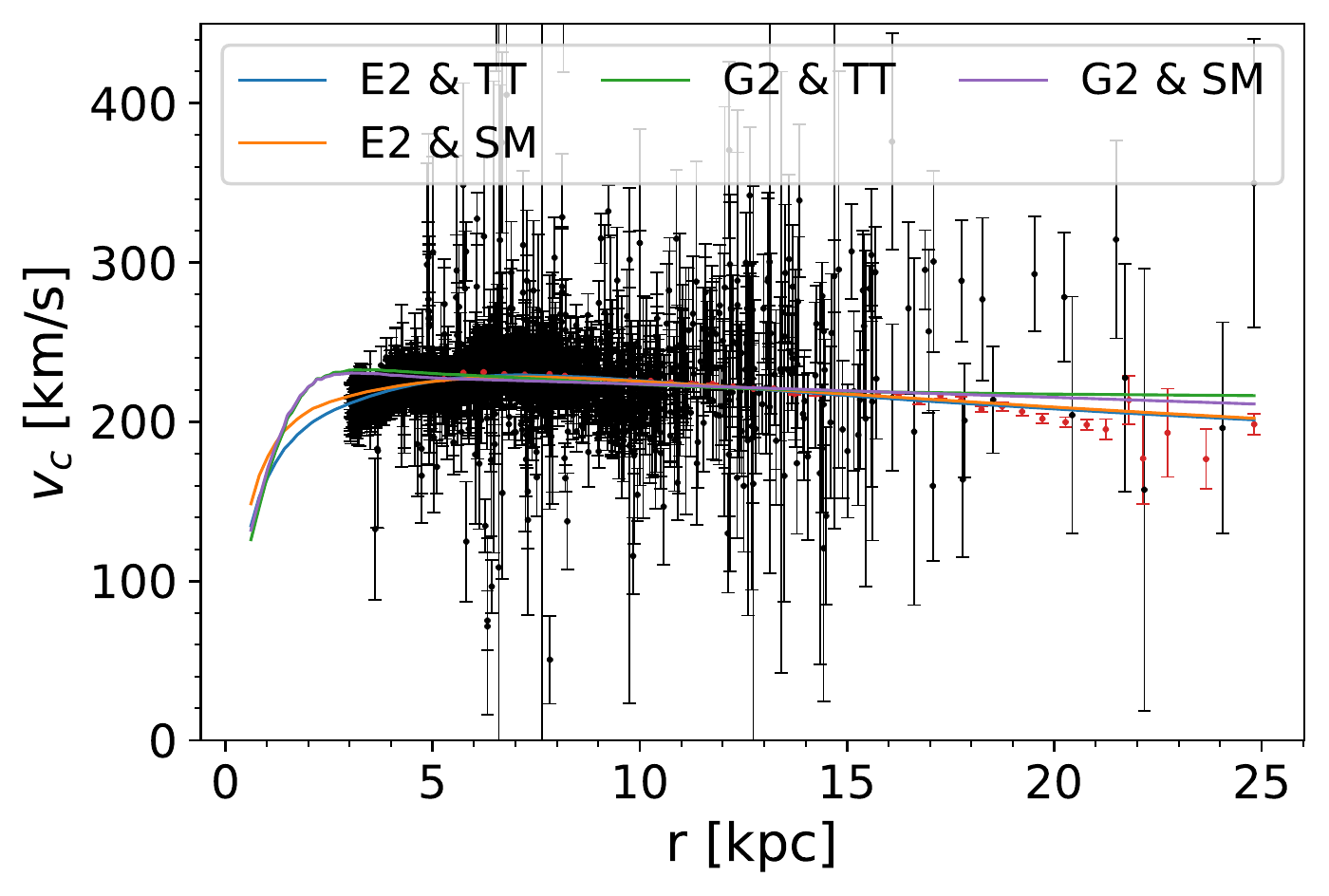} 
\end{subfigure} 
 
\begin{subfigure}{.5\textwidth} 
\centering
\includegraphics[scale=0.6]{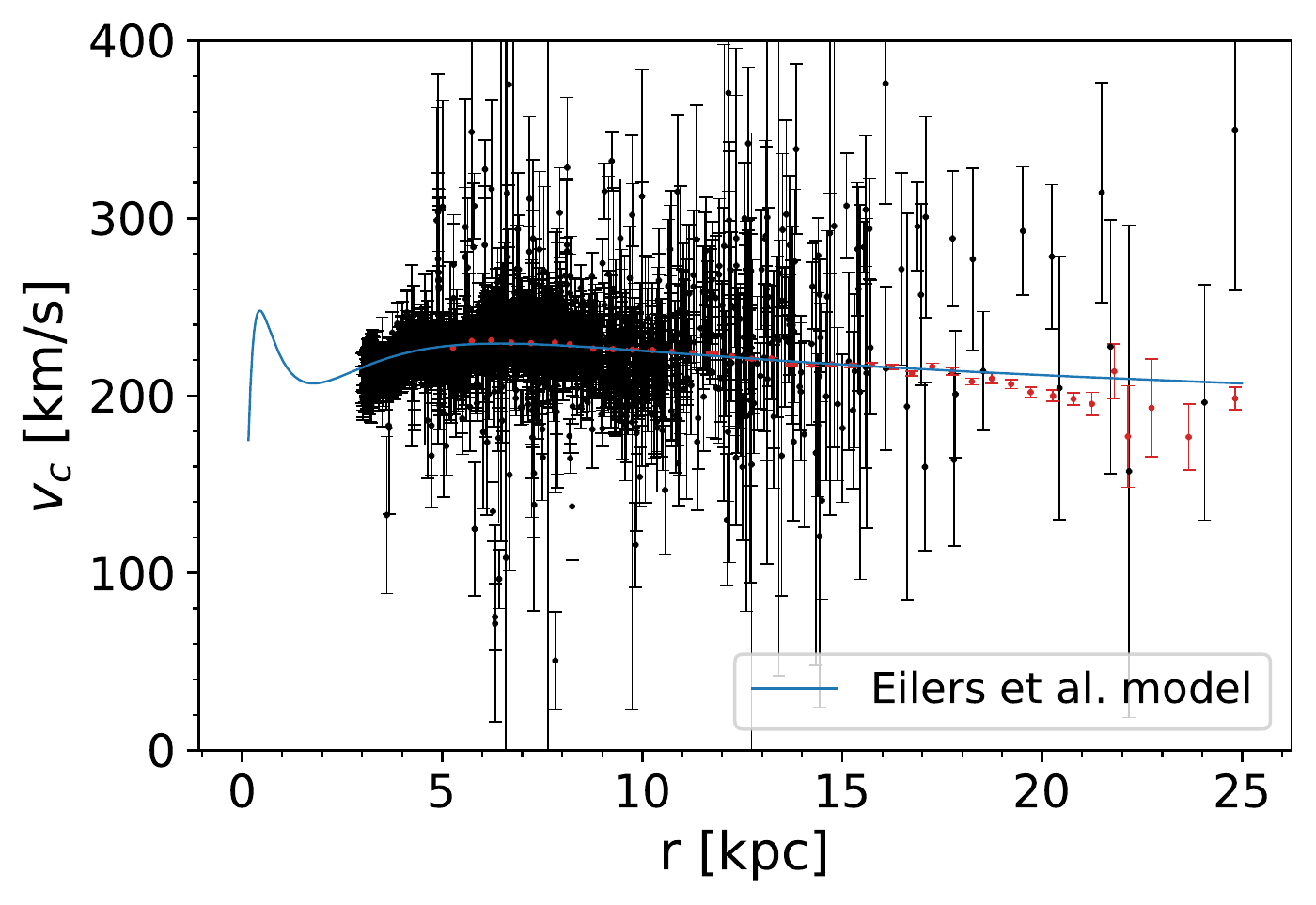}
\end{subfigure} 
\begin{subfigure}{.5\textwidth} 
\centering
\includegraphics[scale=0.6]{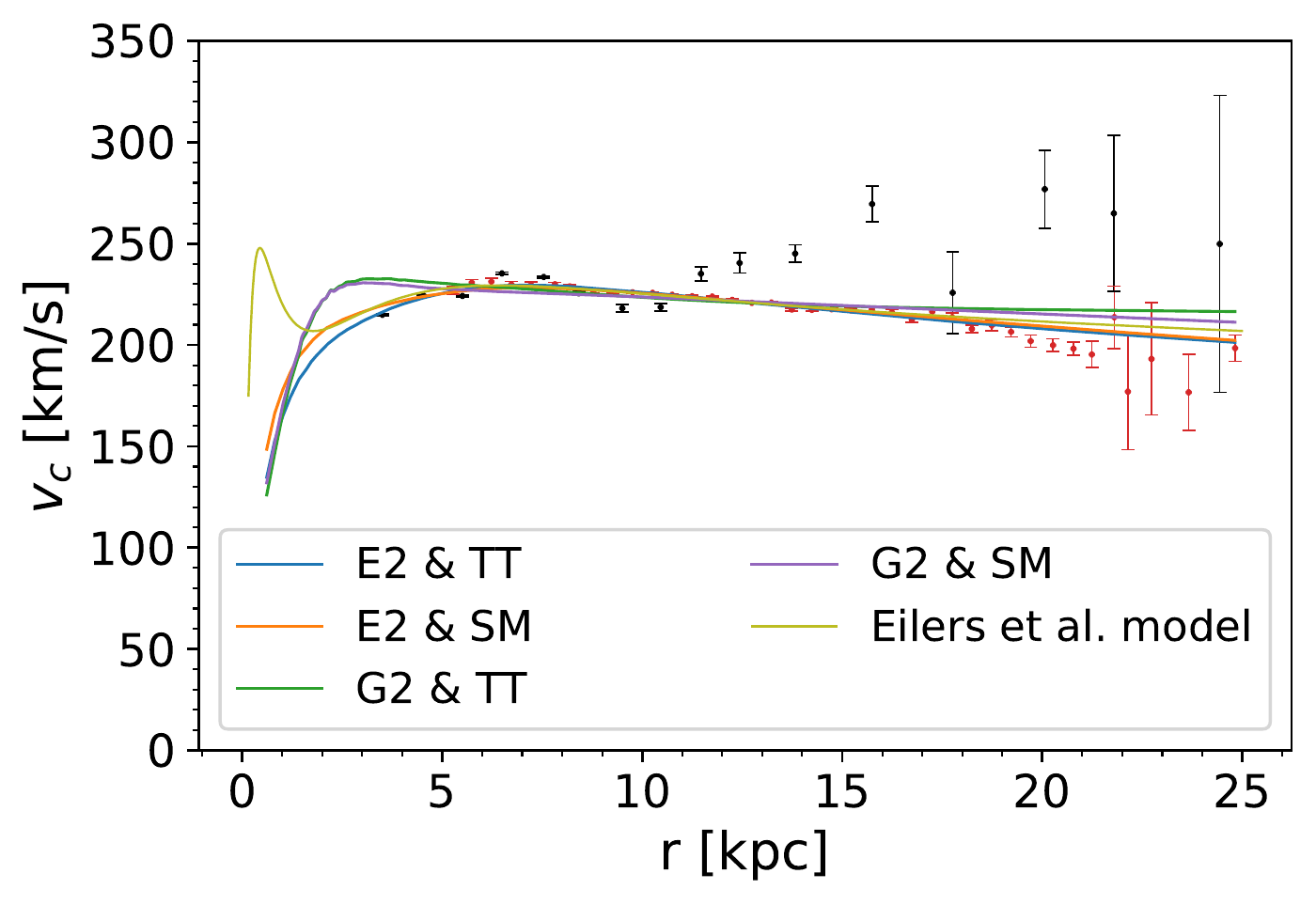}
\end{subfigure} 
\caption{Fits  for the standard gravity case and with a 3 kpc data cut-off. The first plot shows the rotation curve fitted with our four baryonic models and the second plot shows the rotation curve fitted with the Eilers et al. baryonic model. The last plot shows the rotation curve for all 5 baryonic models with the data set {\it galkin} binned.  The black data points are from the {\it galkin} data set and red data points are from \cite{Eilers2018}.}
\label{fig:rotation_curve_stand_grav_cutoff_3}
\end{figure}

\begin{figure}[t]
\centering
\includegraphics[scale=0.6]{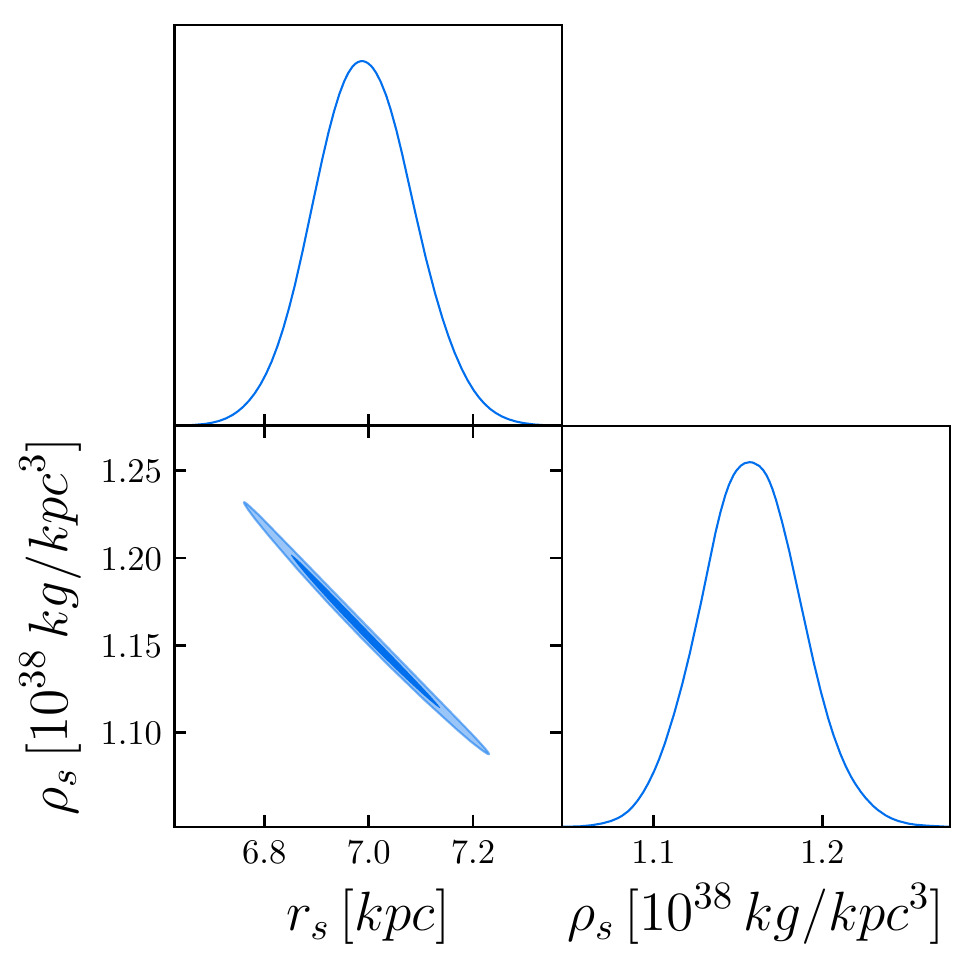}
\caption{Standard gravity case and with a 3 kpc data cut-off . This plot shows, as an example, the confidence regions for the baryonic model E2 \& TT. The dark blue region is the $1 \sigma$ and the light blue region is the $2 \sigma$ confidence region.}
\label{fig:confidence_stand_grav_cutoff_3_E2TT}
\end{figure}

For the case of standard gravity, we only have the two NFW parameters, $r\indice{s}$ and $\rho\indice{s}$ as free fit parameters. The corresponding best-fitting values for the different baryonic models described above, are shown in Table~\ref{tab:results_cut_off_3}. The resulting rotation curves for these values are shown in Fig.~\ref{fig:rotation_curve_stand_grav_cutoff_3}, and, as an example, the confidence regions for the model E2 \& TT are shown in Fig.~\ref{fig:confidence_stand_grav_cutoff_3_E2TT}. The contribution of each component to the rotation curve can be seen in Fig.~\ref{fig:rotation_curve_components_cutoff_3}. As described above, the parameters $r\indice{s}$ and $\rho\indice{s}$ can be recast in terms of the concentration parameter $c$ and the mass of the DM halo $M\indice{200}$. The corresponding values can be read off Table~\ref{tab:results_cutoff_3_mc}.
 
As a test of our machinery but using the Eilers et al. data, we reproduce, with good agreement, the results obtained in \cite{Eilers2018}, where they perform a fit of the rotation data with standard gravity and the baryonic model described in Sec.~\ref{subs:eilers}. The results obtained by \cite{Eilers2018} are also shown in Table~\ref{tab:results_cut_off_3}.

However, these values are not compatible, within the errors, with most of the other results we obtain for standard gravity. This should be due to the different baryonic models used in the analysis. Moreover, we use many more data points for our fits, which can influence the result as well. To check the last point, we fit the Eilers et al. baryonic model to our combined data set with the 3 kpc cut-off. The results are shown in Table~\ref{tab:results_cut_off_3} and \ref{tab:results_cutoff_3_mc} and the corresponding rotation curve is shown in Fig.~\ref{fig:rotation_curve_stand_grav_cutoff_3}. 
The values for the best-fitting parameters remain mostly incompatible with the values found with our baryonic models. We conclude that the reason for this discrepancy should be attributed to the different baryonic models used for our analyses. Interestingly, the best-fit values for $r\indice{s}$ and $\rho\indice{s}$ are now incompatible with the values found by \cite{Eilers2018}. This shows that the dataset {\it galkin} has an influence on the fit too.

For simplicity, in this section we only use the BIC and the $\chi^2\indice{red}$ test as a criteria for model selection. The values for the BIC and $\chi^2\indice{red}$ for all baryonic models are shown in Table~\ref{tab:cut_off_3_modelselection}. By looking at the BIC values for the standard gravity case, it can be seen that the baryonic model E2 \& TT is mostly favoured. 
This is supported by the $\chi^2\indice{red}$ values.

Moreover, the BIC values for the models G2 \& TT and G2 \& SM are significantly higher than the values for the other baryonic models. This could be explained by the baryonic model, which is used for the bulge, G2. The contribution from this bulge model to the rotation curve is very high at small radii as it can be seen in Fig.~\ref{fig:baryonic_components}. Therefore, only a smaller amount of dark matter is allowed at small radii, 
 which results in a worse overall fit.

\begin{table}[t]
\caption{Results for the 3 kpc cut-off data. In this and the following tables, all errors are 68$\%$. The table shows the best-fitting parameter for the different baryonic models. The first entry for each baryonic model is the standard gravity case ($\beta =0$) and the second entry is the modified gravity case ($\beta \neq 0$). For the calculation of $\rho(R_\odot)$ the value for the distance from the sun to the galactic center $R_\odot = (8.122 \pm 0.031)\, \mathrm{kpc}$ \cite{Eilers2018} is used. As a comparison the values from \cite{Eilers2018} for standard gravity, are shown in the last line of this table.}
\label{tab:results_cut_off_3}
\centering
\begin{tabular}{llllll}
\toprule
baryonic model & \(\lambda \, [\mathrm{kpc}]\) & \(\rho\indice{s} \, [10^{38} \mathrm{kg}/\mathrm{kpc}^3]\) & \(r\indice{s}\,[\mathrm{kpc}]\) & \(\beta\)& $\rho(R_\odot)\, [\mathrm{GeV}/\mathrm{cm^3}]$ \\
\midrule
\multirow{2}{*}{E2 \& TT}& & \(1.156\substack{+0.03 \\ -0.028}\) & \(6.99\substack{+0.10 \\ -0.10}\) & 0 & \(0.406\substack{+0.016 \\ -0.016}\)\\
& \(0.133\substack{+0.10 \\ -0.027}\) & \(1.22\substack{+0.07 \\ -0.06}\) & \(6.82\substack{+0.18 \\ -0.18}\) & \(-1.4\substack{+1.7 \\ -5.1}\) & \(0.41\substack{+0.03 \\ -0.03}\)\\
\midrule
\multirow{2}{*}{E2 \& SM} &  & \(4.01\substack{+0.08 \\ -0.08}\) & \(4.30\substack{+0.04 \\ -0.04}\) & 0 & \(0.485\substack{+0.016 \\ -0.015}\) \\
& \(0.224\substack{+0.12 \\ -0.028}\) & \(5.87\substack{+0.26 \\ -0.28}\) & \(3.61\substack{+0.08 \\ -0.07}\) & \(-12.0\substack{+9.1 \\ -1.1}\) & \(0.47\substack{+0.03 \\ -0.03}\)\\
\midrule
\multirow{2}{*}{G2 \& TT} &  & \(0.155\substack{+0.005 \\ -0.005}\) & \(20.4\substack{+0.5 \\ -0.4}\) & 0 & \(0.383\substack{+0.018 \\ -0.018}\)\\
 & \(0.71\substack{+0.11 \\ -0.15}\) & \(0.85\substack{+0.07 \\ -0.05}\) & \(7.85\substack{+0.27 \\ -0.3}\) & \(-6.5\substack{+1.7 \\ -3.7}\) & \(0.38\substack{+0.04 \\ -0.04}\)\\
\midrule
\multirow{2}{*}{G2 \& SM} &  & \(0.996\substack{+0.024 \\ -0.018}\) & \(8.19\substack{+0.09 \\ -0.10}\) & 0 & \(0.483\substack{+0.017 \\ -0.015}\) \\
 & \(0.84\substack{+0.13 \\ -0.13}\) & \(3.88\substack{+0.3 \\ -0.29}\) & \(4.20\substack{+0.15 \\ -0.16}\) & \(-3.2\substack{+0.6 \\ -1.1}\) &\(0.45\substack{+0.05 \\ -0.05}\) \\
\midrule
\multirow{2}{*}{Eilers et al.} &  & \(0.149\substack{+0.006 \\ -0.006}\) & \(18.2\substack{+0.5 \\ -0.5}\) & 0 & \(0.306\substack{+0.018 \\ -0.018}\) \\
& \(0.41\substack{+0.6 \\ -0.05}\) & \(0.301\substack{+0.04 \\ -0.028}\) & \(12.11\substack{+0.7 \\ -0.8}\) & \(-1.8\substack{+1.1 \\ -6.1}\) & \(0.31\substack{+0.05 \\ -0.05}\)\\
(taken from \cite{Eilers2018}) &  & \(0.211 \pm 0.018\) & \(14.8\pm 0.4\) & 0 & \(0.30\pm 0.03\) \\
\bottomrule
\end{tabular}
\end{table}

\begin{table}[t]
\caption{Results for the 3 kpc cut-off   data. The table shows the best-fitting parameter for the different baryonic models. The first entry for each baryonic model is the standard gravity case ($\beta =0$) and the second entry is the modified gravity case ($\beta \neq 0$). The best-fitting parameters are written in terms of the concentration parameter and virial mass of the dark matter halo. As a comparison the values from \cite{Eilers2018} for standard gravity, are shown in the last line of this table.}
\label{tab:results_cutoff_3_mc}
\centering
\begin{tabular}{lllll}
\toprule
baryonic model & \(\lambda\, [\mathrm{kpc}]\) & c & \(M\indice{200}\, (\ten{11}\,M_\odot)\) & \(\beta\)\\
\midrule
\multirow{2}{*}{E2 \& TT} &  & \(23.92\substack{+0.24 \\ -0.23}\) & \(5.63\substack{+0.28 \\ -0.29}\) & 0 \\
& \(0.133\substack{+0.10 \\ -0.027}\) & \(24.4\substack{+0.5 \\ -0.5}\) & \(5.5\substack{+0.6 \\ -0.5}\) & \(-1.4\substack{+1.7 \\ -5.1}\)\\
\midrule
\multirow{2}{*}{E2 \& SM} &  & \(38.43\substack{+0.29 \\ -0.29}\) & \(5.43\substack{+0.21 \\ -0.19}\) & 0\\
 & \(0.224\substack{+0.12 \\ -0.028}\) & \(44.4\substack{+0.7 \\ -0.8}\) & \(4.9\substack{+0.4 \\ -0.4}\) & \(-12.0\substack{+9.1 \\ -1.1}\)\\
\midrule
\multirow{2}{*}{G2 \& TT} &  & \(10.82\substack{+0.14 \\ -0.14}\) & \(13.04\substack{+1.0 \\ -1.0}\) & 0 \\
& \(0.71\substack{+0.11 \\ -0.15}\) & \(21.2\substack{+0.7 \\ -0.5}\) & \(5.6\substack{+0.8 \\ -0.8}\) & \(-6.5\substack{+1.7 \\ -3.7}\)\\
\midrule
\multirow{2}{*}{G2 \& SM} &  & \(22.57\substack{+0.21 \\ -0.16}\) & \(7.6\substack{+0.3 \\ -0.3}\) & 0\\
& \(0.84\substack{+0.13 \\ -0.13}\) & \(38.0\substack{+1.3 \\ -1.1}\) & \(4.9\substack{+0.7 \\ -0.7}\) & \(-3.2\substack{+0.6 \\ -1.1}\)\\
\midrule
\multirow{2}{*}{Eilers et al.} &  & \(11.44\substack{+0.21 \\ -0.21}\) & \(7.8\substack{+0.8 \\ -0.8}\) & 0\\
& \(0.41\substack{+0.6 \\ -0.05}\) & \(15.4\substack{+0.8 \\ -0.7}\) & \(5.5\substack{+1.4 \\ -1.3}\) & \(-1.8\substack{+1.1 \\ -6.1}\)\\
taken from \cite{Eilers2018} &  & \(12.8 \pm 0.3\) & \(7.25\pm 0.25\) & 0 \\
\bottomrule
\end{tabular}
\end{table}

\begin{table}[t]
\caption{Results for the 3 kpc cut-off   data. The table shows the BIC and $\chi^2\indice{red}$ values for all baryonic models.  For the differences between the BICs of the standard and modified gravity case, we report the absolute value.}
\label{tab:cut_off_3_modelselection}
\centering
\begin{tabular}{lllll}
\toprule
baryonic model & & BIC & $\Delta \mathrm{BIC}$ & $\chi^2\indice{red}$\\
\midrule
\multirow{2}{*}{E2 \& TT} & stand. grav. & 3445.5 & \multirow{2}{*}{17.9} & 1.31\\
& mod. grav. & 3463.4 & & 1.31\\
\midrule
\multirow{2}{*}{E2 \& SM} & stand. grav. & 3527.3 & \multirow{2}{*}{70.35} & 1.34 \\
& mod. grav. & 3456.95 & & 1.31 \\
\midrule
\multirow{2}{*}{G2 \& TT} & stand. grav. & 4802.2 & \multirow{2}{*}{1229.7} & 1.82\\
& mod. grav. & 3572.5 & & 1.35\\
\midrule
\multirow{2}{*}{G2 \& SM} & stand. grav.& 4550.1 & \multirow{2}{*}{1069.3} & 1.73\\
& mod. grav. & 3480.8 & & 1.31 \\
\midrule
\multirow{2}{*}{Eilers et al.} & stand. grav. & 3569.0 & \multirow{2}{*}{429.99} & 1.35\\
& mod.grav. & 3998.99 & & 1.51\\
\bottomrule
\end{tabular}
\end{table}

\subsubsection{Modified gravity}
\begin{figure}[t]
\begin{subfigure}{.5\textwidth} 
\centering
\includegraphics[scale=0.6]{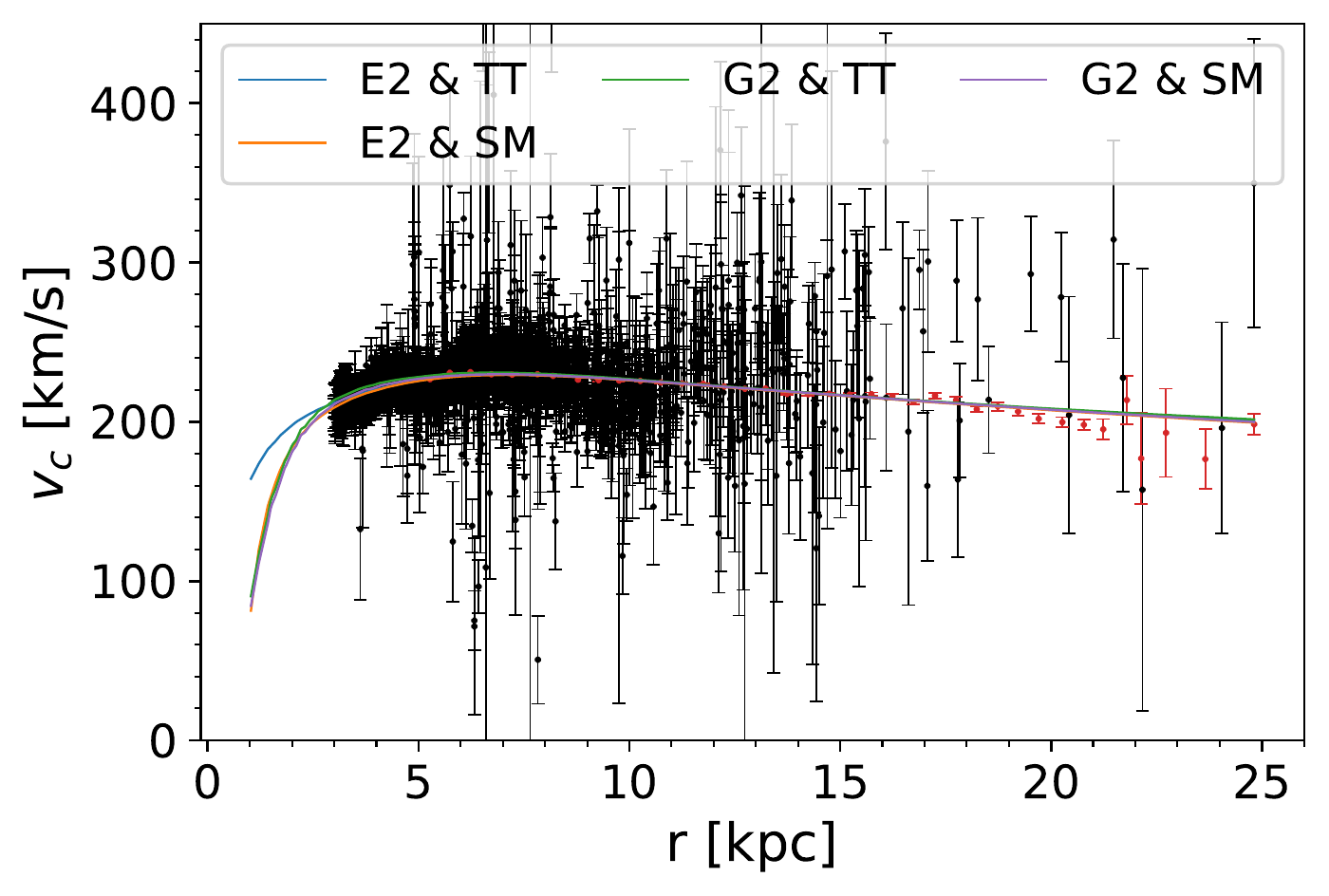}
\end{subfigure} 
 
\begin{subfigure}{.5\textwidth} 
\centering
\includegraphics[scale=0.6]{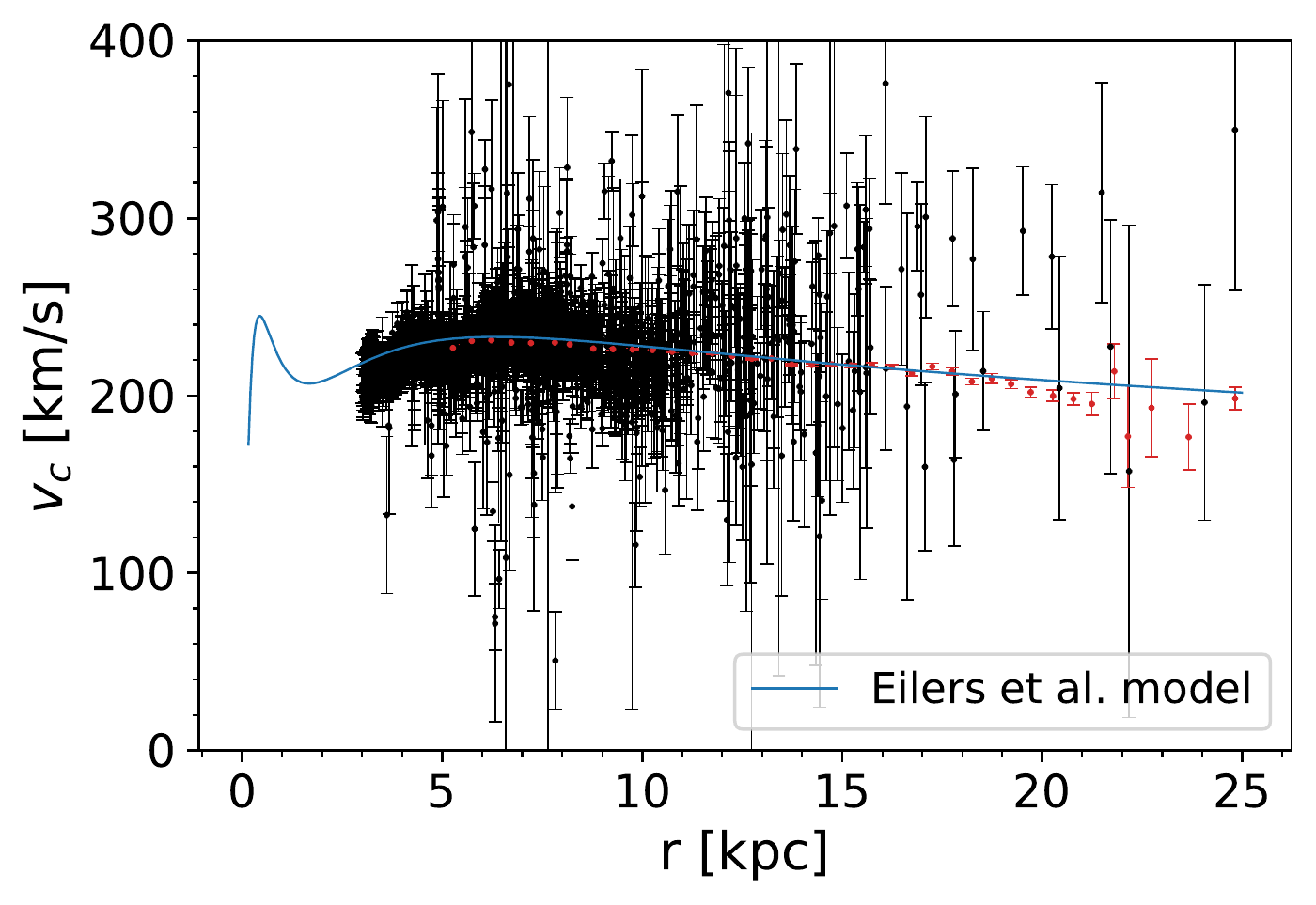} 
\end{subfigure} 
\begin{subfigure}{.5\textwidth} 
\centering
\includegraphics[scale=0.6]{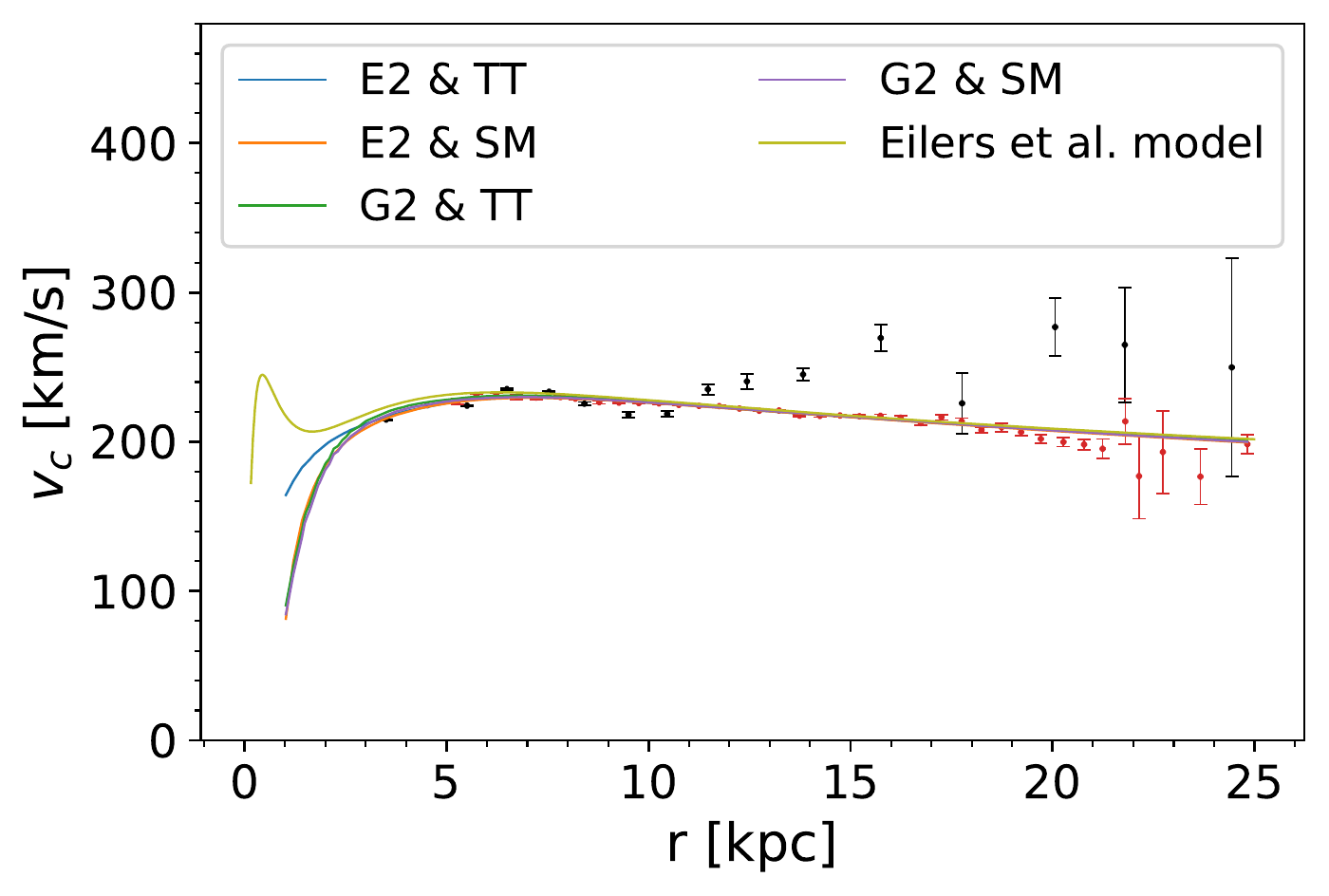} 
\end{subfigure} 
\caption{Results for modified gravity and 3 kpc data cut-off. The first plot shows the rotation curve fitted with our four baryonic models and the second plot shows the rotation curve fitted with the Eilers et al. baryonic model. The last plot shows the rotation curve for all 5 baryonic models with the data set {\it galkin} binned.  The black data points are from the {\it galkin} data set and red data points are from \cite{Eilers2018}.}
\label{fig:rotation_curve_mod_grav_cutoff_3}
\end{figure}

\begin{figure}[t]
\centering
\includegraphics[scale=0.6]{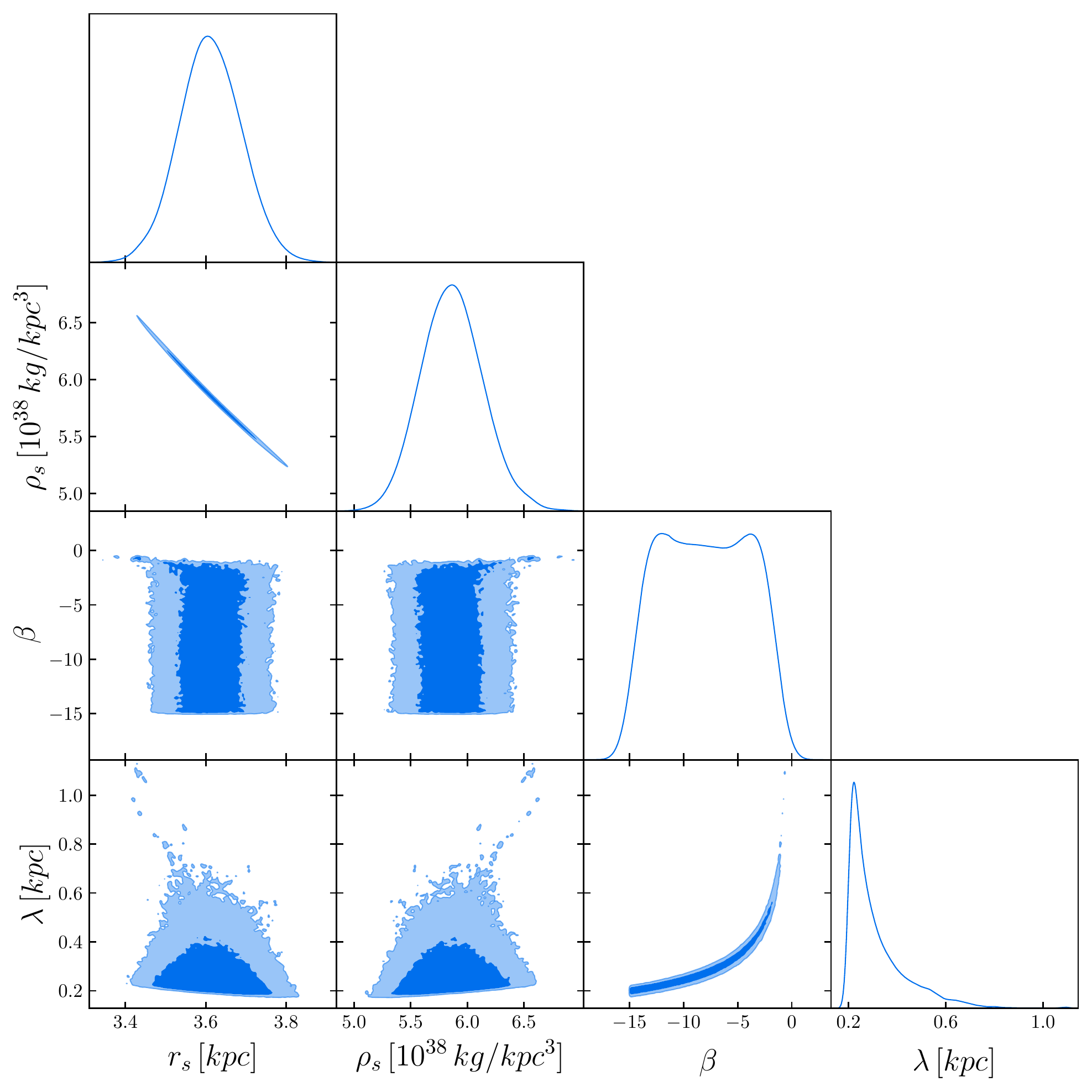}
\caption{Results for the 3 kpc data cut-off. This plot shows, as an example, the confidence regions for the baryonic model E2 \& SM and modified gravity. The dark blue region is the $1 \sigma$ and the light blue region is the $2 \sigma$ confidence region.}
\label{fig:confidence_mod_grav_cutoff_3_E2SM}
\end{figure}

\begin{figure}[t]
\centering
\includegraphics[scale=0.6]{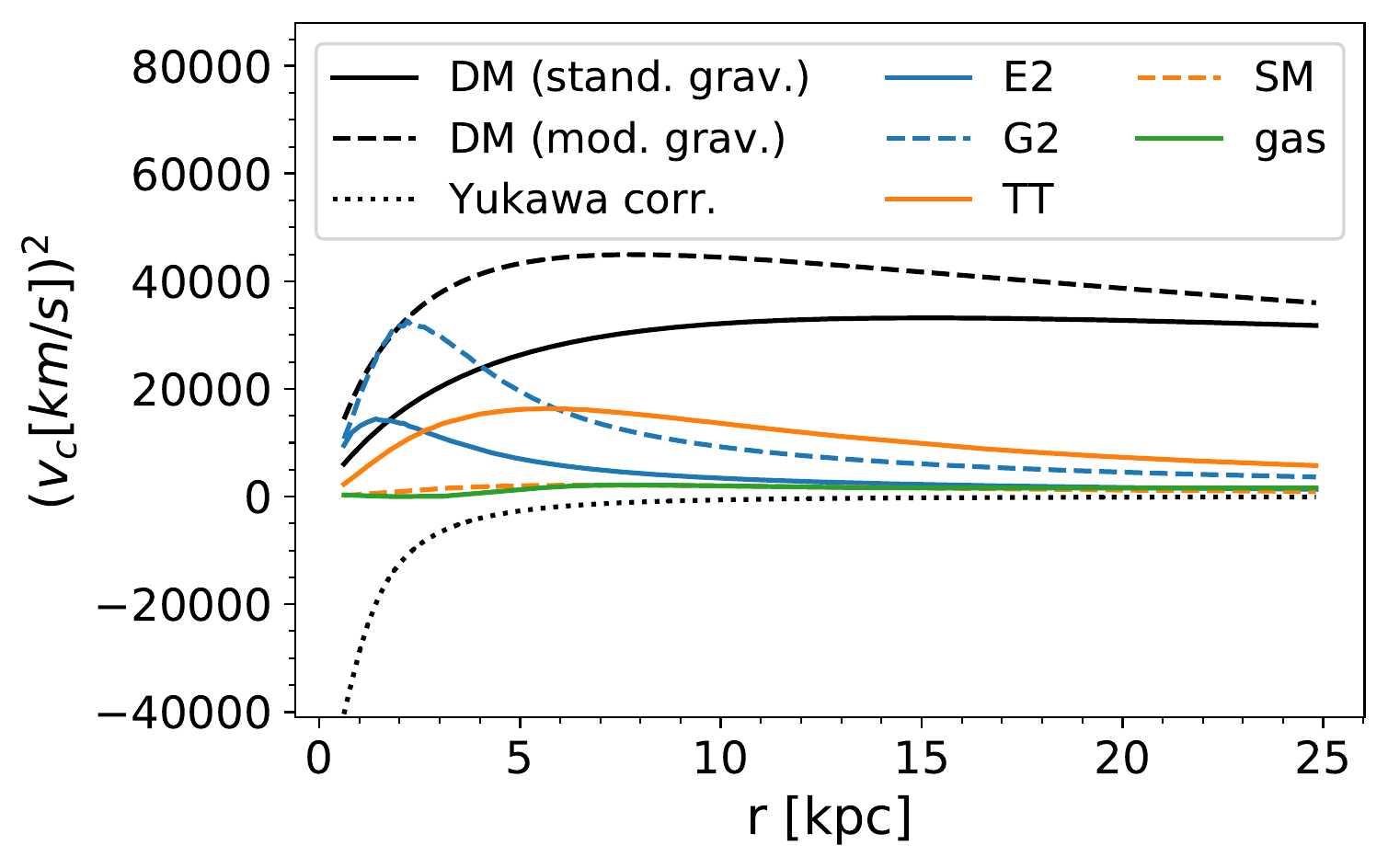}
\caption{Results for the 3 kpc data cut-off. This plot shows the contribution of all different components to the rotation curve. For the standard and modified gravity contribution, only the result for the best fitting case is shown.}
\label{fig:rotation_curve_components_cutoff_3}
\end{figure}

For the case of modified gravity, we have two more free parameters, the Yukawa strength $\beta$ and the Yukawa length $\lambda$. The best-fitting parameters for the modified gravity case are shown in Table~\ref{tab:results_cut_off_3}, while the corresponding rotation curves are shown in Fig.~\ref{fig:rotation_curve_mod_grav_cutoff_3}. 
As an example, the confidence regions for the baryonic model E2 \& SM are shown in Fig.~\ref{fig:confidence_mod_grav_cutoff_3_E2SM}. The contribution of each component to the rotation curve can be seen in Fig.~\ref{fig:rotation_curve_components_cutoff_3}.

The results show that the Yukawa strength $\beta$ is negative for all baryonic models. However, due to the large errors, some of the values for $\beta$, are compatible with zero, i.e. standard gravity. 

The values for the Yukawa scaling length $\lambda$ are smaller than 1 kpc for all baryonic models. This means that the Yukawa correction is mainly dominant in the inner regions of the galaxy. At \emph{our} distance from the galaxy center ($R\simeq 8 \mathrm{kpc}$), the Yukawa strength reduces to $\sim (10^{-27}- 10^{-5})\cdot\beta$ depending on the baryonic model. Furthermore, the fact that the Yukawa correction is repulsive (i.e. negative value of $\beta$) means that the Yukawa correction reduces the contribution of dark matter to the total rotation curve in the very inner regions.

The uncertainties on $\beta$ and $\lambda$ are very large. This is mainly due to a degeneracy between the two parameters since, for small values of $\lambda$, the modiﬁed gravity part signiﬁcantly changes the rotation curve in the region where we excluded the data. This shows that data at small radii would be needed to better constrain the modiﬁed gravity parameters, so an approach not based on the circular approximation would be called for.

It is interesting to notice that the baryonic model E2 \& TT provides values of $\beta$, and, hence, of $\rho\indice{s}$ and $r\indice{s}$, compatible with the standard gravity case.

Expressing the two parameters of the NFW proﬁle in terms of c and $M_{200}$, we see that the amount of dark matter which is needed to describe the rotation curve is compatible within the errors for all baryonic models. The corresponding values are shown in Table~\ref{tab:results_cutoff_3_mc}.
Comparing these values with the results of the standard gravity case, we ﬁnd that with the Yukawa correction less dark matter is needed. As expected, this is true for all baryonic models. 
Depending on the chosen baryonic model, the amount of dark matter needed is reduced by roughly 10 \% - 57 \%. The model E2 \& TT is excluded from this consideration because the NFW parameters for this model are so similar to the standard gravity case.

\begin{figure}[t]
\centering
\includegraphics[scale=0.5]{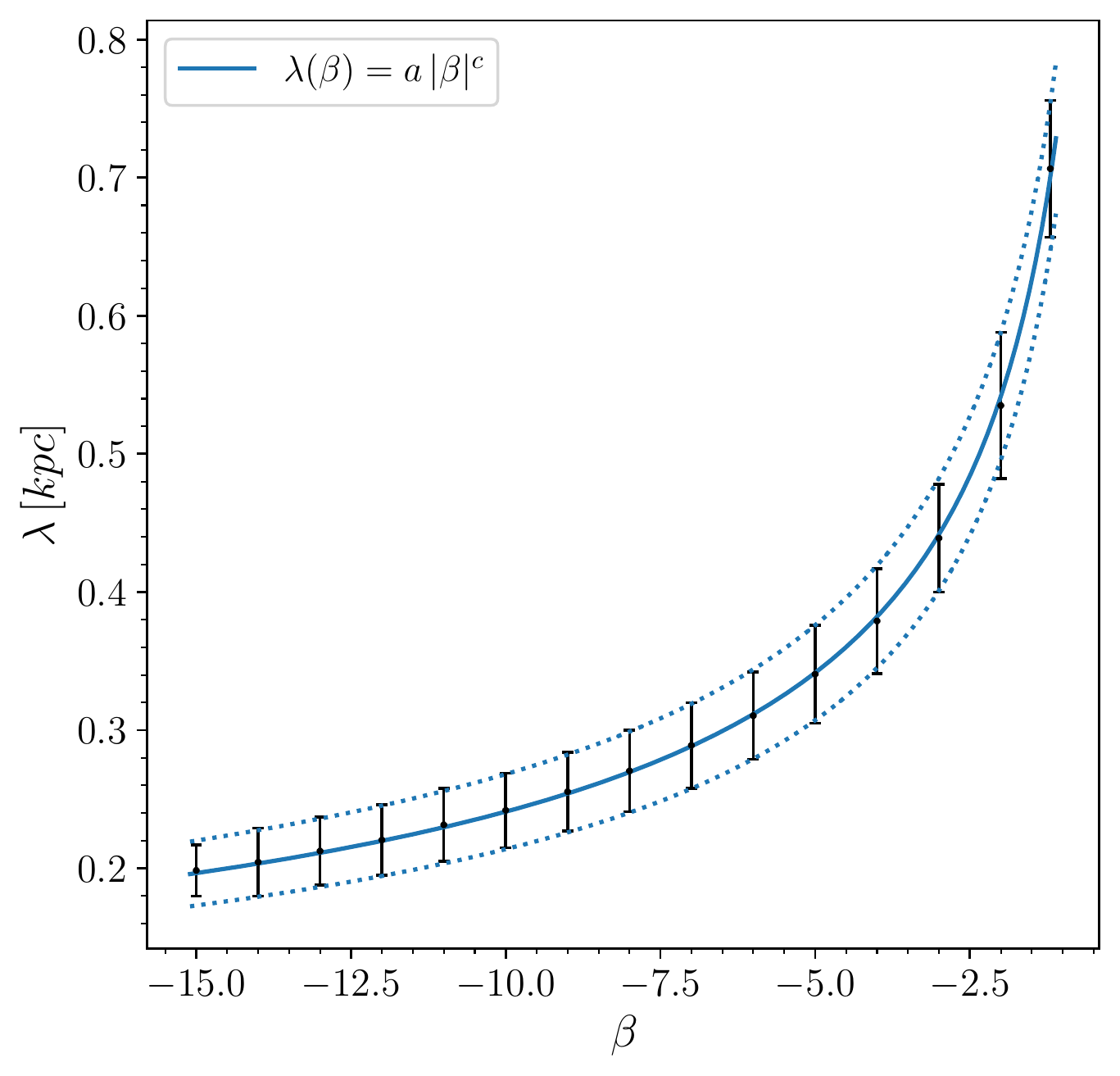}
\caption{The plot shows the polynomial fit which describes the degeneracy between $\lambda$ and $\beta$.}
\label{fig:lambda(beta)}
\end{figure}

As mentioned before, there is a clear degeneracy between $\beta$ and $\lambda$ (see e.g Fig.~\ref{fig:confidence_mod_grav_cutoff_3_E2SM}). The $\lambda (\beta)$ function can be better constrained than the individual parameters. In the following we give a first estimation of this function. To do so we take the possible range of values for $\lambda$ for fixed values of $\beta$ from Fig.~\ref{fig:confidence_mod_grav_cutoff_3_E2SM}. These values are plotted in Fig.~\ref{fig:lambda(beta)} and are used for a fit with a polynomial function. The used fit function is 
\begin{align}
\lambda(\beta)=a\,|\beta|^c   \label{eq:fitlambda}
\end{align}
and the resulting values for the fit parameters are $a=0.77\pm 0.06\,\mathrm{kpc}$ and $c=-0.503^{+0.016}_{-0.019}$. The resulting fit is also shown in Fig.~\ref{fig:lambda(beta)}.

For simplicity, we again only use the BIC and a $\chi^2\indice{red}$ test as a model selection criteria for this section.
The corresponding values for the different baryonic models are shown in Table~\ref{tab:cut_off_3_modelselection}. 
Looking at the BIC values for the modified gravity cases, the baryonic model E2 \& SM seems to be the most favoured. 
Moreover, comparing the BIC values for the standard gravity cases with the modified gravity ones, the G2 \& TT and G2 \& SM plus modified gravity are strongly favoured compared to the standard gravity case. This can be explained with the baryonic model of the bulge, G2. As stated above, the model G2 has a higher contribution for small radii and, hence, the standard gravity case is not able to describe the data well at such distances from the galactic center.  
The repulsive Yukawa correction reduces the additional contribution from the dark matter at small radii in such a way that the rotation curve is ﬁtted better.
Overall, the baryonic model E2 \& TT for the standard gravity case is the most favoured.

In conclusion, we argue that standard gravity is enough to describe the data with a radius larger than 3 kpc: as it can be seen in section \ref{subsection:full_data_set} the modiﬁed gravity parameters are mostly constrained by the questionable data below 3 kpc.

\subsection{Cut-off at 5 kpc}
\begin{figure}[t]
\begin{subfigure}{.5\textwidth} 
\centering
\includegraphics[scale=0.6]{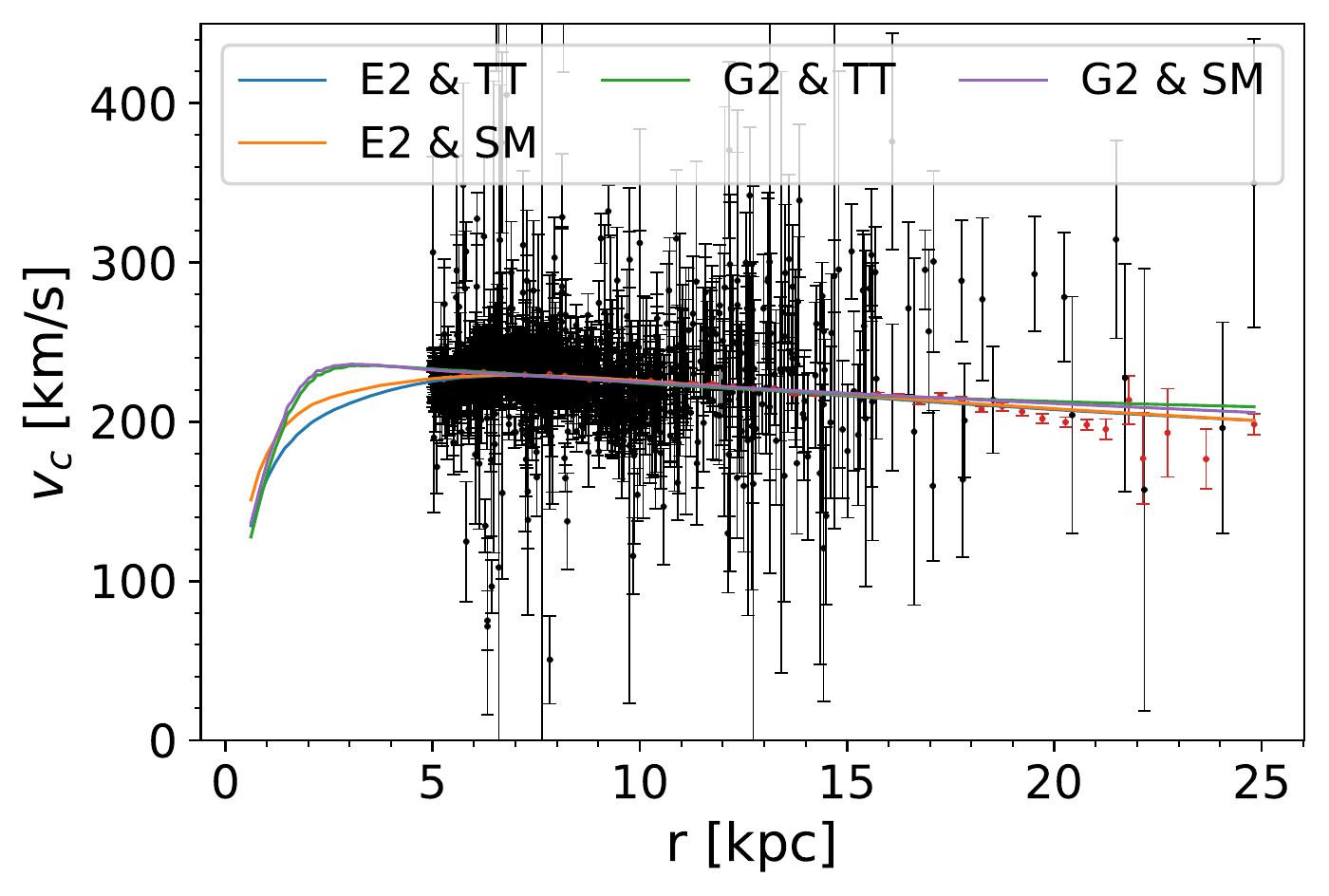} 
\end{subfigure} 
 
\begin{subfigure}{.5\textwidth} 
\centering
\includegraphics[scale=0.6]{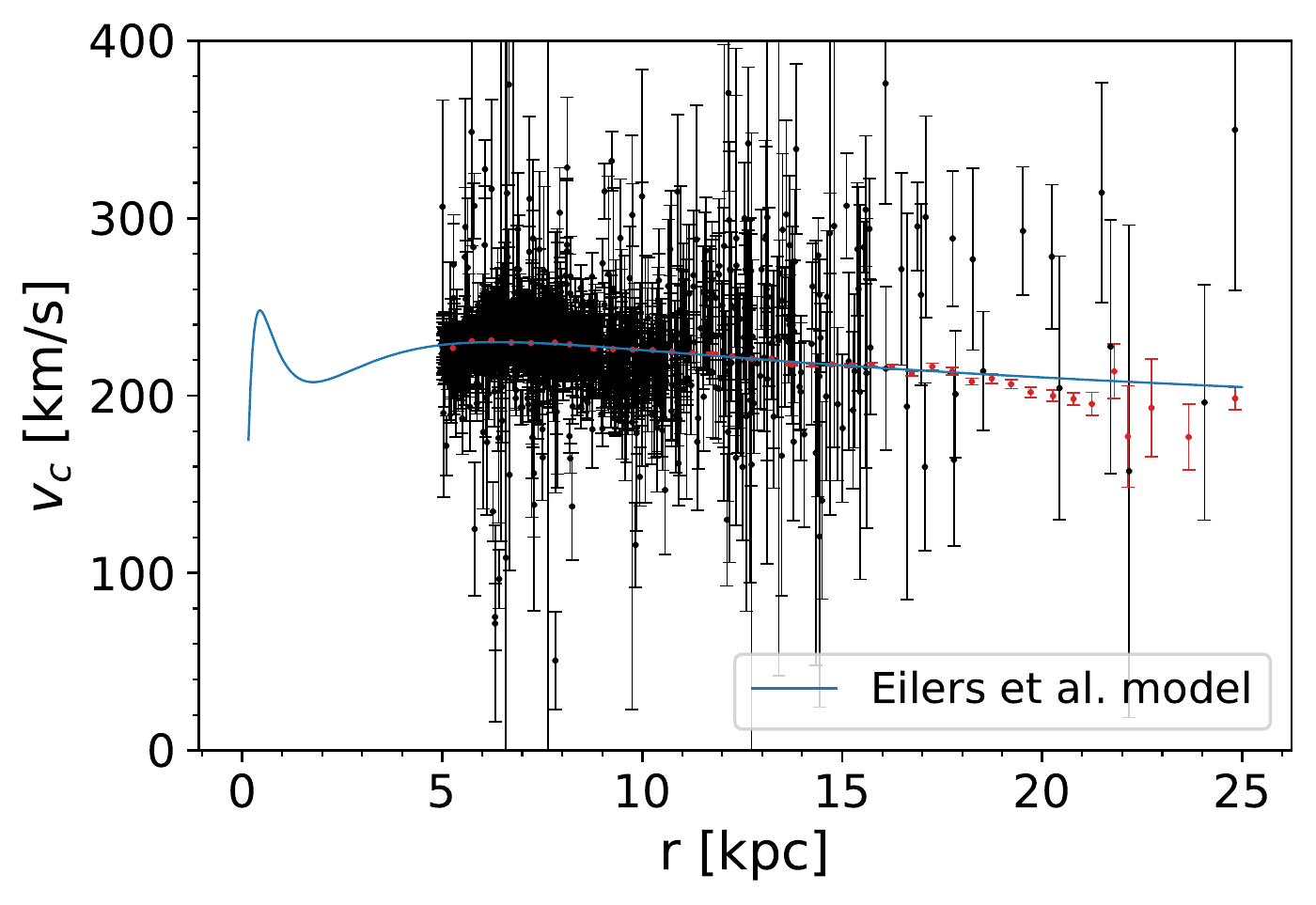} 
\end{subfigure}
\begin{subfigure}{.5\textwidth} 
\centering
\includegraphics[scale=0.6]{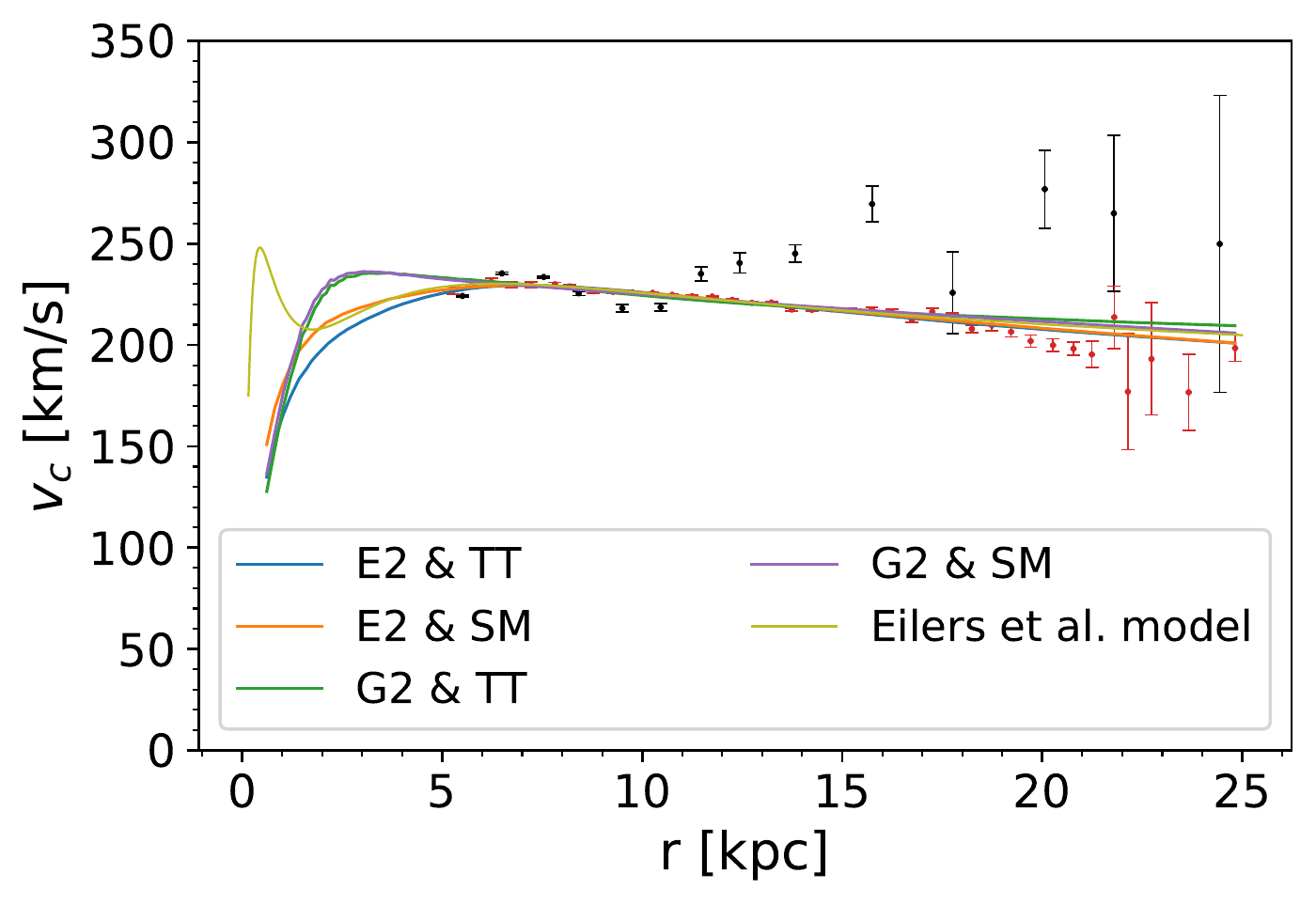}
\end{subfigure}
\caption{Results for standard gravity and  5 kpc data cut-off. The first plot shows the rotation curve fitted with our four baryonic models and the second plot shows the rotation curve fitted with the Eilers et al. baryonic model. The last plot shows the rotation curve for all 5 baryonic models with the data set {\it galkin} binned.  The black data points are from the {\it galkin} data set and red data points are from \cite{Eilers2018}.}
\label{fig:rotation_curve_stand_grav_cutoff_5}
\end{figure}

\begin{figure}[t]
\begin{subfigure}{.5\textwidth} 
\centering
\includegraphics[scale=0.6]{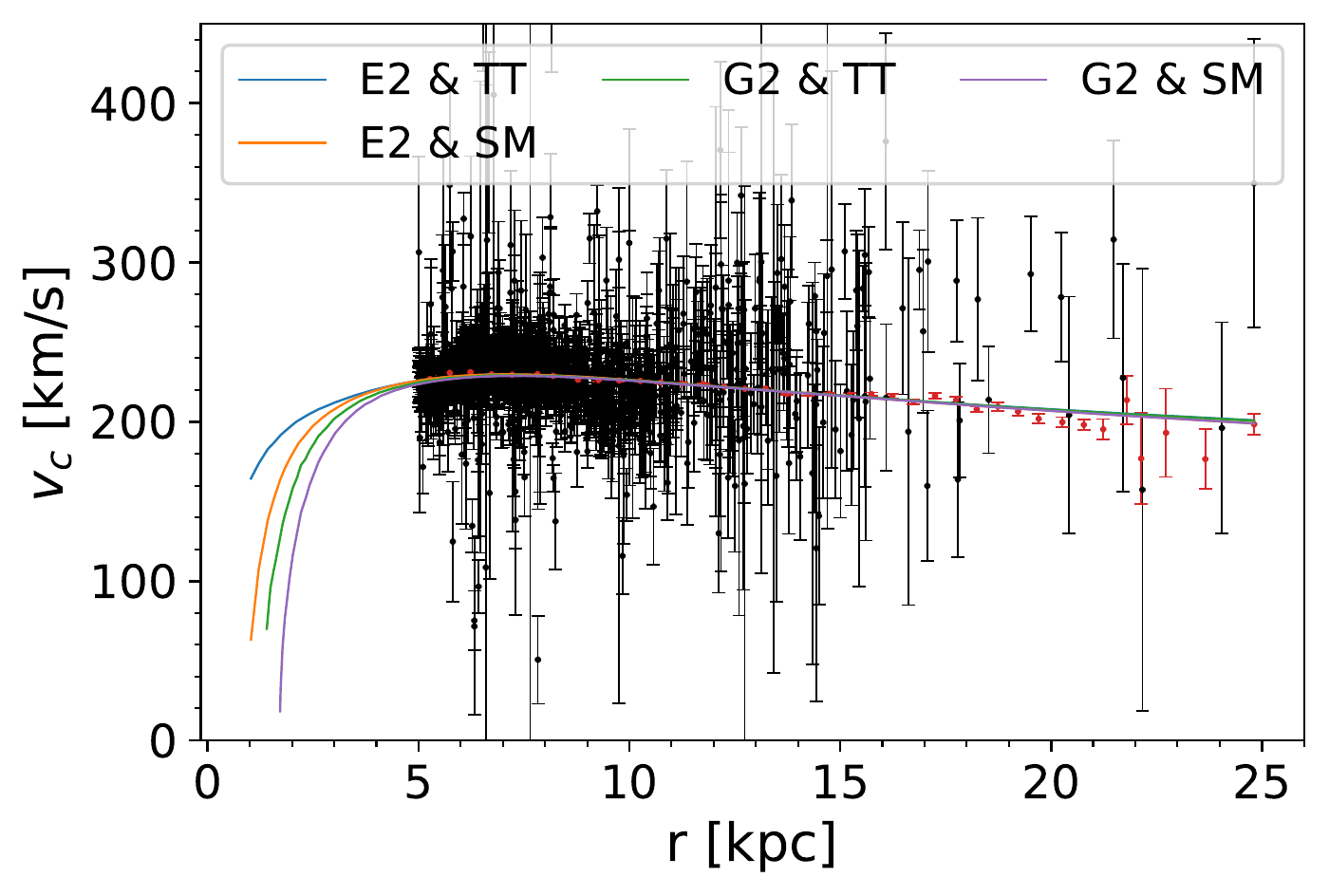} 
\end{subfigure} 
\begin{subfigure}{.5\textwidth} 
\centering
\includegraphics[scale=0.6]{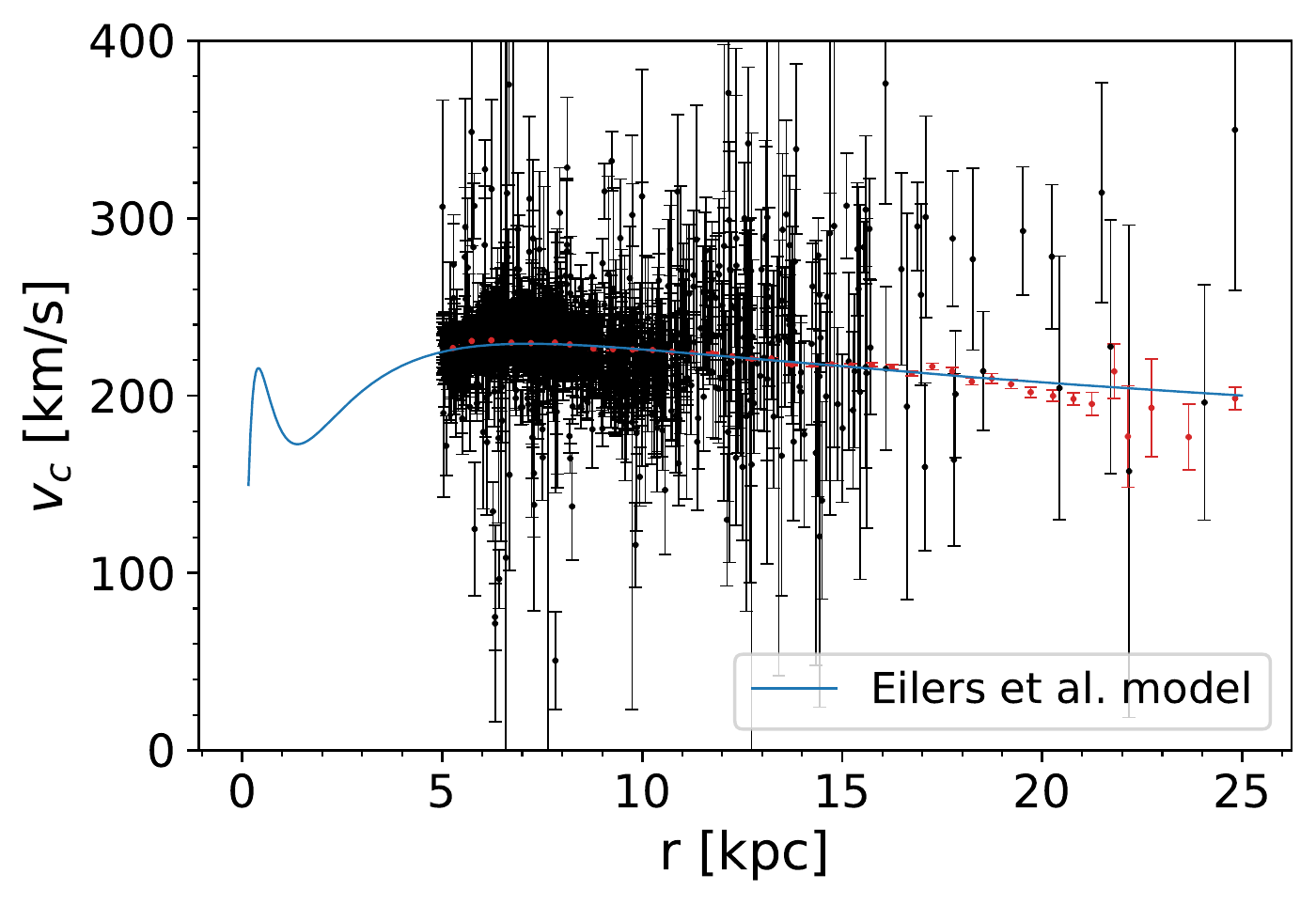} 
\end{subfigure}
\begin{subfigure}{.5\textwidth} 
\centering
\includegraphics[scale=0.6]{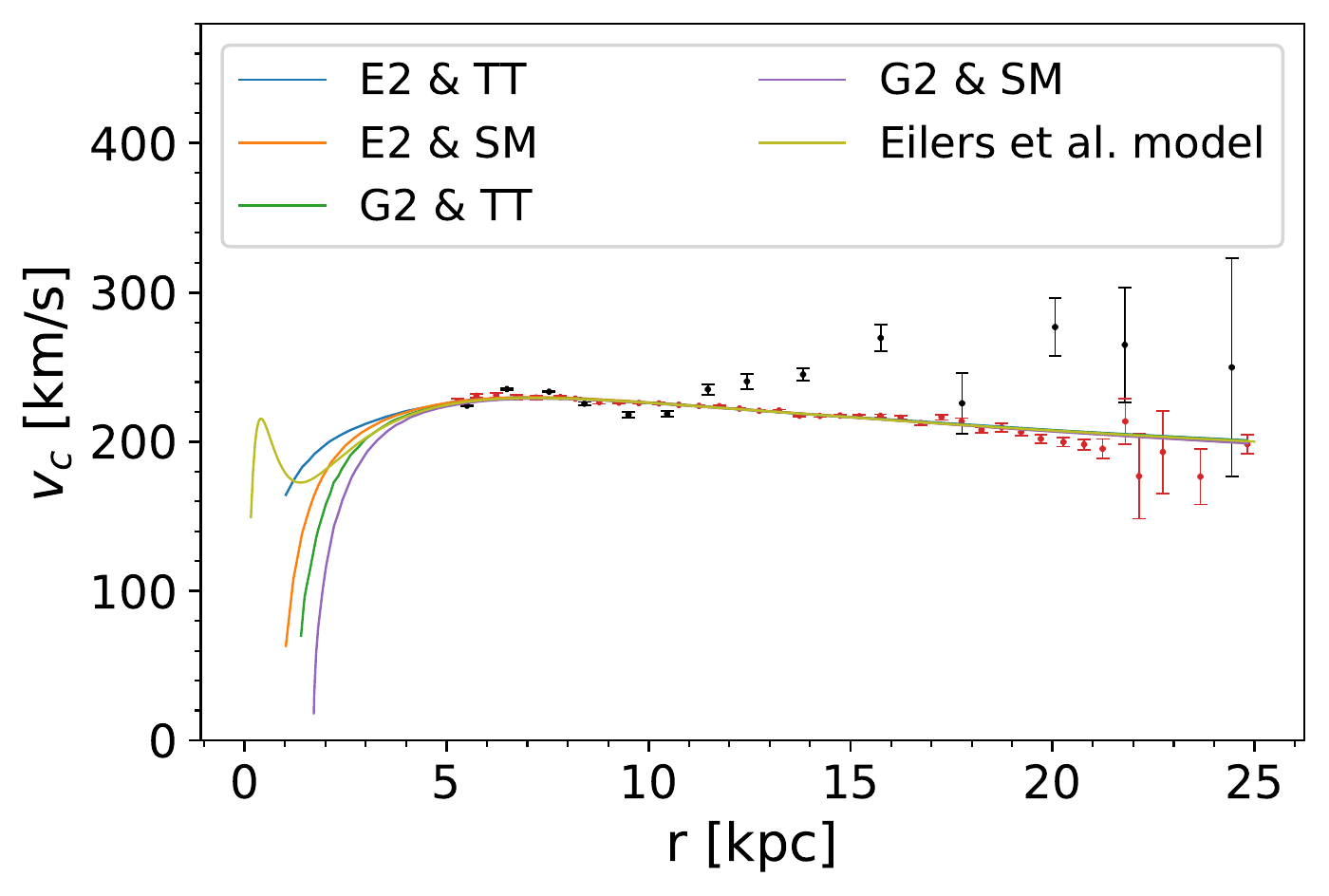} 
\end{subfigure}
\caption{Results for modified gravity and  5 kpc data cut-off. The first plot shows the rotation curve fitted with our four baryonic models and the second plot shows the rotation curve fitted with the Eilers et al. baryonic model. The last plot shows the rotation curve for all 5 baryonic models with the data set {\it galkin} binned.  The black data points are from the {\it galkin} data set and red data points are from \cite{Eilers2018}.}
\label{fig:rotation_curve_mod_grav_cutoff_5}
\end{figure}

\begin{figure}[t]
\centering
\includegraphics[scale=0.6]{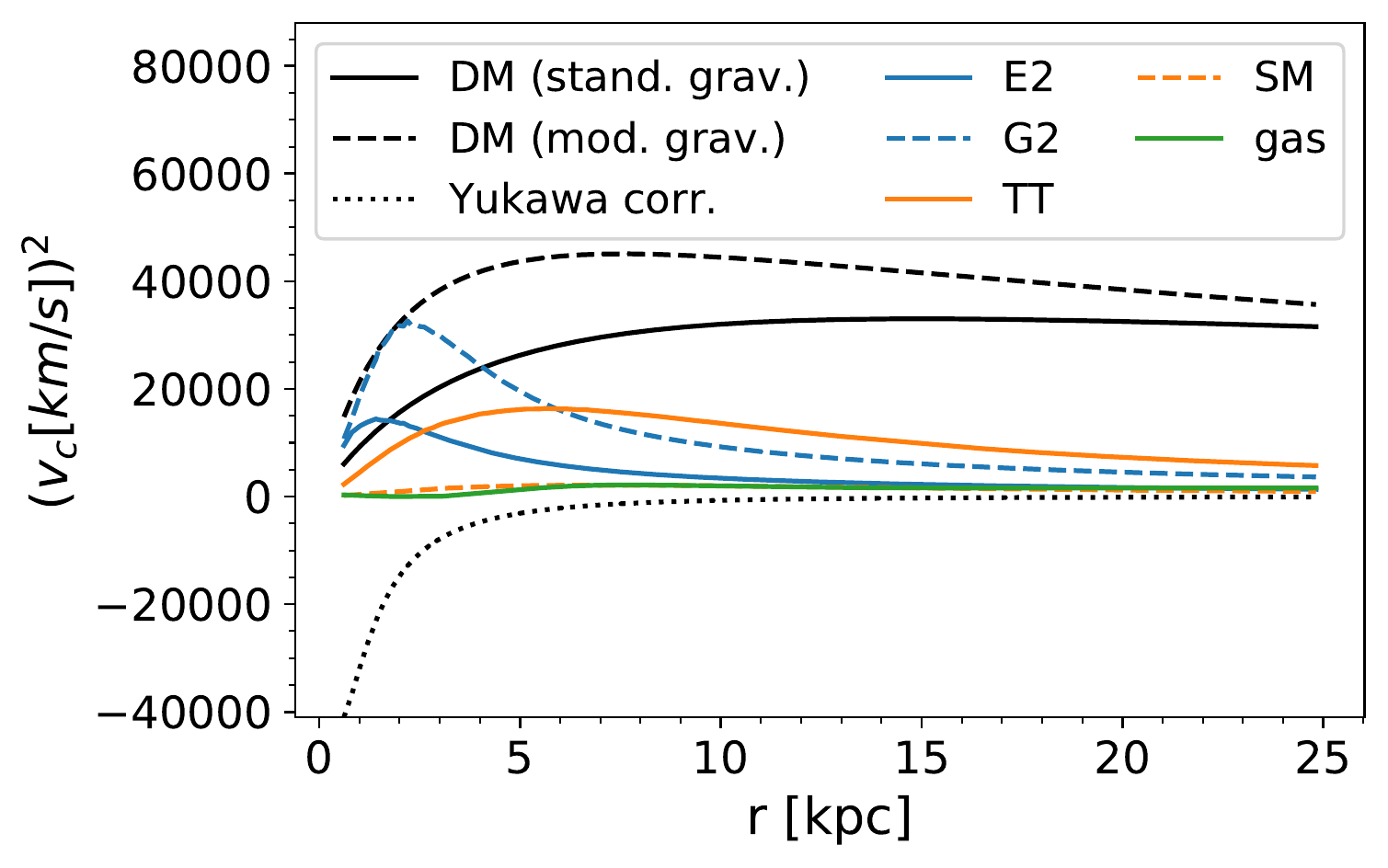}
\caption{Results for the 5 kpc data cut-off. This plot shows the contribution of all different components to the rotation curve. For the standard and modified gravity contribution, only the result for the best fitting case is shown.}
\label{fig:rotation_curve_components_cutoff_5}
\end{figure}

In addition to the analysis with the 3 kpc cut-off, we also perform an analysis with a 5 kpc cut-off.  
This cut-off is chosen because the far more precise data of \cite{Eilers2018} starts at $\sim 5\, \mathrm{kpc}$. As before, the analysis is carried out for all the five baryonic models and for both the standard and the modified gravity case. 
The best-fitting values are shown in Table~\ref{tab:results_cutoff_5} and the corresponding rotation curves are shown in Fig.~\ref{fig:rotation_curve_stand_grav_cutoff_5} and Fig.~\ref{fig:rotation_curve_mod_grav_cutoff_5}.
The contribution of each component to the rotation curve can be seen in Fig.~\ref{fig:rotation_curve_components_cutoff_5}.
 
The results obtained in the modified gravity case are not deviating significantly from the analysis with the 3 kpc cut-off. 
This holds for all baryonic models, due to the large errors especially for $\beta$ and $\lambda$. 

For the standard gravity case, some of the best-fitting values for the 5 kpc cut-off are deviating significantly from the 3 kpc cut-off values, but the deduced conclusions are unchanged.
For simplicity, only the BIC and the $\chi^2\indice{red}$ test are used for model selection, as before. The corresponding values are shown in Table~\ref{tab:cut_off_5_modelselection}. 
A comparison of the different values of the BIC shows, that, in the modified gravity case, all baryonic models, except the model G2 \& SM, fit equally well. 
Definitely, we find with strong evidence that the standard gravity case of the model E2 \& TT is the most favoured.

\begin{table}[t]
\caption{Results for the  5 kpc data cut-off. The table shows the best-fitting parameter for the different baryonic models. The first entry for each baryonic model is the standard gravity case ($\beta =0$) and the second entry is the modified gravity case ($\beta \neq 0$). For the calculation of $\rho(R_\odot)$ the value for the distance from the sun to the galactic center $R_\odot = (8.122 \pm 0.031)\, \mathrm{kpc}$ \cite{Eilers2018} is used.}
\label{tab:results_cutoff_5}
\centering
\begin{tabular}{llllll}
\toprule
baryonic model & \(\lambda\, [\mathrm{kpc}]\) & \(\rho\indice{s}\, [10^{38} \mathrm{kg}/\mathrm{kpc}^3]\) & \(r\indice{s}\,[\mathrm{kpc}]\) & \(\beta\) & $\rho(R_\odot)\, [\mathrm{GeV}/\mathrm{cm^3}]$\\
\midrule
\multirow{2}{*}{E2 \& TT} &  & \(1.17\substack{+0.03 \\ -0.04}\) & \(6.93\substack{+0.14 \\ -0.10}\) & 0 & \(0.405\substack{+0.021 \\ -0.019}\)\\
& \(0.129\substack{+0.12 \\ -0.023}\) & \(1.22\substack{+0.10 \\ -0.09}\) & \(6.81\substack{+0.28 \\ -0.26}\) & \(-1.3\substack{+3.1 \\ -9.3}\) & \(0.41\substack{+0.05 \\ -0.05}\)\\
\midrule
\multirow{2}{*}{E2 \& SM} &  & \(4.57\substack{+0.13 \\ -0.12}\) & \(4.04\substack{+0.05 \\ -0.05}\) & 0 & \(0.478\substack{+0.020 \\ -0.020}\) \\
 & \(0.28\substack{+0.14 \\ -0.06}\) & \(6.3\substack{+0.7 \\ -0.6}\) & \(3.49\substack{+0.15 \\ -0.16}\) & \(-8.9\substack{+4.9 \\ -4.1}\) & \(0.47\substack{+0.07 \\ -0.07}\) \\
\midrule
\multirow{2}{*}{G2 \& TT} &  & \(0.250\substack{+0.011 \\ -0.007}\) & \(15.1\substack{+0.3 \\ -0.3}\) & 0 & \(0.374\substack{+0.022 \\ -0.018}\)\\
& \(0.55\substack{+0.12 \\ -0.03}\) & \(0.95\substack{+0.10 \\ -0.08}\) & \(7.42\substack{+0.3 \\ -0.4}\) & \(-13.5\substack{+4.8 \\ -0.9}\) & \(0.38\substack{+0.05 \\ -0.05}\)\\
\midrule
\multirow{2}{*}{G2 \& SM} &  & \(1.53\substack{+0.04 \\ -0.04}\) & \(6.53\substack{+0.09 \\ -0.10}\) & 0 & \(0.466\substack{+0.019 \\ -0.019}\)\\
& \(0.46\substack{+0.12 \\ -0.03}\) & \(4.4\substack{+0.4 \\ -0.3}\) & \(3.97\substack{+0.15 \\ -0.16}\) & \(-13.1\substack{+5.6 \\ -1.0}\) & \(0.44\substack{+0.06 \\ -0.06}\) \\
\midrule
\multirow{2}{*}{Eilers et al.} &  & \(0.178\substack{+0.008 \\ -0.008}\) & \(16.3\substack{+0.5 \\ -0.5}\) & 0 & \(0.304\substack{+0.021 \\ -0.020}\)\\
 & \(0.46\substack{+0.17 \\ -0.06}\) & \(0.37\substack{+0.05 \\ -0.04}\) & \(10.8\substack{+0.7 \\ -0.7}\) & \(-12.2\substack{+6.5 \\ -1.3}\) & \(0.31\substack{+0.05 \\ -0.05}\) \\
\bottomrule
\end{tabular}
\end{table}

\begin{table}[t]
\caption{Results for the  5 kpc data cut-off. The table shows the BIC and $\chi^2\indice{red}$ values for all baryonic models.  For the differences between the BICs of the standard and modified gravity case, we report the absolute value.}
\label{tab:cut_off_5_modelselection}
\centering
\begin{tabular}{lllll}
\toprule
baryonic model & & BIC & $\Delta \mathrm{BIC}$ & $\chi^2\indice{red}$\\
\midrule
\multirow{2}{*}{E2 \& TT}& stand. grav. & 3247.7 & \multirow{2}{*}{11.8} & 1.60\\
& mod. grav & 3259.5 & & 1.60\\
\midrule
\multirow{2}{*}{E2 \& SM}& stand. grav. & 3257.5 & \multirow{2}{*}{1.9} & 1.61\\
& mod. grav. & 3259.4 & & 1.60\\
\midrule
\multirow{2}{*}{G2 \& TT}& stand. grav.& 3492.6 & \multirow{2}{*}{232.9} & 1.72\\
& mod. grav.& 3259.7 & & 1.60\\
\midrule
\multirow{2}{*}{G2 \& SM} & stand. grav.& 3409.99 & \multirow{2}{*}{140.03} & 1.73\\
& mod. grav.& 3269.96 & & 1.61\\
\midrule
\multirow{2}{*}{Eilers et al.} & stand. grav.& 3288.2 & \multirow{2}{*}{29.1} & 1.62 \\
& mod. grav.& 3259.1 & & 1.60\\
\bottomrule
\end{tabular}
\end{table}

\subsection{Full data set}\label{subsection:full_data_set}

In addition to the previous cases, where we have excluded the innermost region of the data, we now perform for completeness the analysis with the full dataset. Nevertheless, it is worth underlining that the results should be carefully interpreted because of the intrinsic tension shown by the data at small radii.
As before, this section is divided into two parts describing the standard and the modified gravity cases, respectively.

\subsubsection{Standard gravity}
\begin{figure}[t]
\begin{subfigure}{.5\textwidth} 
\centering
\includegraphics[scale=0.6]{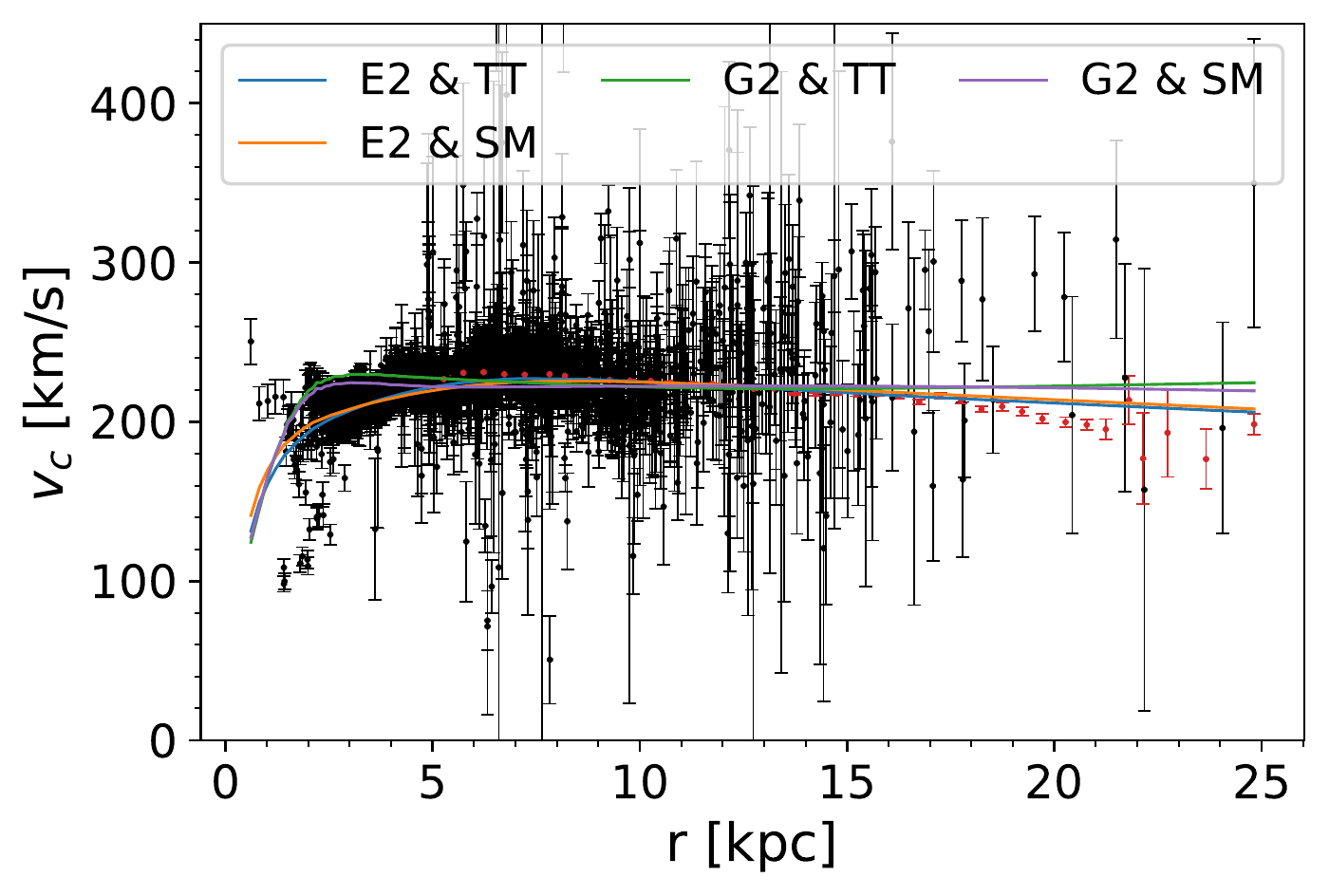} 
\end{subfigure} 

\begin{subfigure}{.5\textwidth} 
\centering
\includegraphics[scale=0.6]{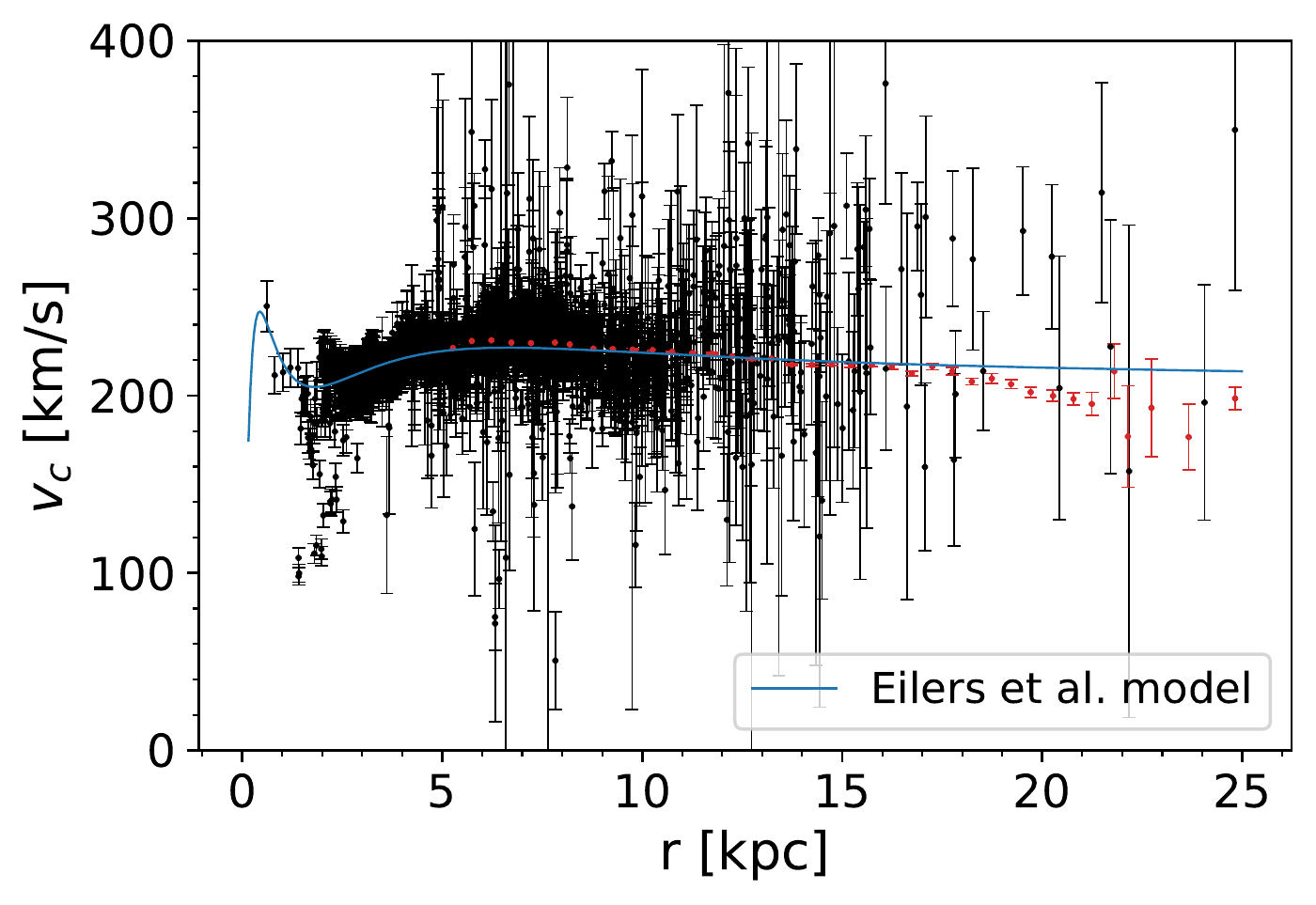}
\end{subfigure}
\begin{subfigure}{.5\textwidth} 
\centering
\includegraphics[scale=0.6]{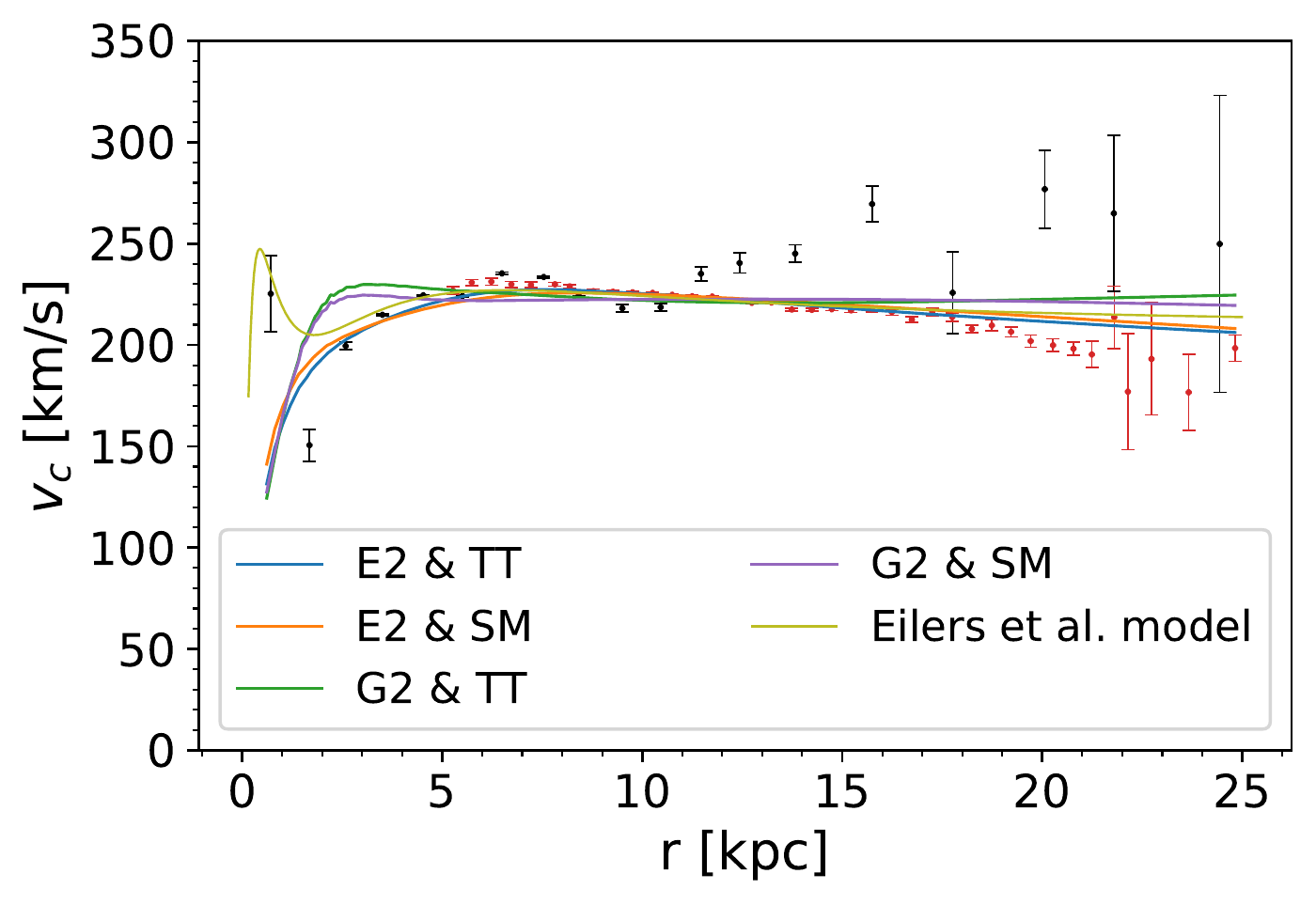}
\end{subfigure}
\caption{Standard gravity and full data set (no cut-off). The first plot shows the rotation curve fitted with our four baryonic models and the second plot shows the rotation curve fitted with the Eilers et al. baryonic model. The last plot shows the rotation curve for all 5 baryonic models with the data set {\it galkin} binned.  The black data points are from the {\it galkin} data set and red data points are from \cite{Eilers2018}.}
\label{fig:rotation_curve_stand_grav_all_data}
\end{figure}

The values of the best-fitting parameters are shown in Table~\ref{tab:results_all_data} and the corresponding rotation curves are shown in Fig.~\ref{fig:rotation_curve_stand_grav_all_data}. In Fig. \ref{fig:rotation_curve_components_fulldata} the contribution of each component to the rotation curve is shown. As before, we calculate from the NFW parameters the concentration parameter and the virial mass of the dark matter halo. The corresponding values are shown in Table~\ref{tab:results_all_data_mc}.

For model selection, we used the BIC, the $\chi^2\indice{red}$ test and the Bayes ratio. In this analysis, the Likelihoods of the different parameters have a shape closer to a Gaussian and thus the use of the Bayes ratio for model selection seems appropriate. 
The values for the $\chi^2\indice{red}$ and the BIC are shown in Table~ \ref{tab:all_data_modelselection} and the values for the Bayes ratio are shown in Table~ \ref{tab:bayesratio_stand_grav}. 

Using the Bayes ratio, 
 we find that the model E2 \& TT is the most favoured one. Moreover, using the Kass-Raftery scale, it is evident that the E2 \& TT model is strongly favoured than all the others. This is supported by looking at the BIC values for the different baryonic models. 
Also the $\chi^2\indice{red}$ value for the model E2 \& TT is the smallest and therefore the $\chi^2\indice{red}$ test supports the findings of the Bayes ratio.

\begin{table}[t]
\caption{Results for the full data set (no cut-off). The table shows the best-fitting parameter for the different baryonic models. The first entry for each baryonic model is the standard gravity case ($\beta =0$) and the second entry is the modified gravity case ($\beta \neq 0$). For the calculation of $\rho(R_\odot)$ the value for the distance from the sun to the galactic center $R_\odot = (8.122 \pm 0.031)\, \mathrm{kpc}$ \cite{Eilers2018} is used.}
\label{tab:results_all_data}
\centering

\begin{tabular}{llllll}
\toprule
baryonic model & \(\lambda \,[\mathrm{kpc}]\) & \(\rho\indice{s} \,[10^{38} \mathrm{kg}/\mathrm{kpc}^3]\) & \(r\indice{s}\,[\mathrm{kpc}]\) & \(\beta\) & $\rho(R_\odot)\, [\mathrm{GeV}/\mathrm{cm^3}]$ \\
\midrule
\multirow{2}{*}{E2 \& TT} & & \(0.787\substack{+0.018 \\ -0.017}\) & \(8.61\substack{+0.11 \\ -0.11}\) & 0 & \(0.422\substack{+0.014 \\ -0.014}\)\\
 & \(0.391\substack{+0.021 \\ -0.023}\) & \(2.23\substack{+0.08 \\ -0.07}\) & \(5.06\substack{+0.09 \\ -0.09}\) & \(-7.7\substack{+0.7 \\ -0.9}\) & \(0.391\substack{+0.021 \\ -0.020}\) \\
\midrule
\multirow{2}{*}{E2 \& SM}& & \(2.54\substack{+0.04 \\ -0.04}\) & \(5.39\substack{+0.04 \\ -0.05}\) & 0 & \(0.513\substack{+0.013 \\ -0.013}\) \\
& \(0.517\substack{+0.024 \\ -0.021}\) & \(7.94\substack{+0.25 \\ -0.27}\) & \(3.15\substack{+0.05 \\ -0.05}\) & \(-3.70\substack{+0.23 \\ -0.26}\) & \(0.458\substack{+0.023 \\ -0.023}\)\\
\midrule
\multirow{2}{*}{G2 \& TT}& & \(0.0876\substack{+0.0029 \\ -0.0029}\) & \(30.2\substack{+0.8 \\ -0.7}\) & 0 & \(0.387\substack{+0.019 \\ -0.018}\) \\
& \(0.726\substack{+0.024 \\ -0.027}\) & \(1.02\substack{+0.06 \\ -0.05}\) & \(7.16\substack{+0.17 \\ -0.21}\) & \(-7.6\substack{+0.4 \\ -0.5}\) & \(0.377\substack{+0.028 \\ -0.029}\) \\
\midrule
\multirow{2}{*}{G2 \& SM} & & \(0.582\substack{+0.011 \\ -0.010}\) & \(11.07\substack{+0.12 \\ -0.13}\) & 0 & \(0.505\substack{+0.014 \\ -0.015}\)\\
& \(0.842\substack{+0.028 \\ -0.029}\) & \(4.48\substack{+0.25 \\ -0.21}\) & \(3.94\substack{+0.09 \\ -0.11}\) & \(-3.57\substack{+0.16 \\ -0.18}\) & \(0.44\substack{+0.03 \\ -0.04}\)  \\
\midrule
\multirow{2}{*}{Eilers et al.}& & \(0.085\substack{+0.004 \\ -0.003}\) & \(26.6\substack{+0.7 \\ -0.8}\) & 0 & \(0.311\substack{+0.019 \\ -0.018}\)\\ 
& \(0.281\substack{+0.016 \\ -0.017}\) & \(0.675\substack{+0.026 \\ -0.028}\) & \(7.82\substack{+0.18 \\ -0.17}\) & \(-38.1\substack{+3.9 \\ -5.1}\) & \(0.299\substack{+0.018 \\ -0.018}\)\\ 
\bottomrule
\end{tabular}
\end{table}

\begin{table}[t]
\caption{Results for the full data set (no cut-off). The table shows the best-fitting parameter for the different baryonic models. The first entry for each baryonic model is the standard gravity case ($\beta =0$) and the second entry is the modified gravity case ($\beta \neq 0$). The best-fitting parameters are written in terms of the concentration parameter and virial mass of the dark matter halo.}
\label{tab:results_all_data_mc}
\centering

\begin{tabular}{lllll}
\toprule
baryonic model & \(\lambda [\mathrm{kpc}]\) & \(c\) & \(M\indice{200}\, (\ten{11}\,M_\odot)\) & \(\beta\) \\
\midrule
\multirow{2}{*}{E2 \& TT} & & \(20.60\substack{+0.18 \\ -0.18}\) & \(6.7\substack{+0.3 \\ -0.3}\) \\
& \(0.391\substack{+0.021 \\ -0.023}\) & \(30.8\substack{+0.4 \\ -0.4}\) & \(4.55\substack{+0.30 \\ -0.29}\) & \(-7.7\substack{+0.7 \\ -0.9}\)\\
\midrule
\multirow{2}{*}{E2 \& SM} & & \(32.34\substack{+0.20 \\ -0.19}\) & \(6.39\substack{+0.20 \\ -0.20}\) \\
& \(0.517\substack{+0.024 \\ -0.021}\) & \(49.7\substack{+0.6 \\ -0.6}\) & \(4.61\substack{+0.27 \\ -0.27}\) & \(-3.70\substack{+0.23 \\ -0.26}\)\\
\midrule
\multirow{2}{*}{G2 \& TT} & & \(8.55\substack{+0.12 \\ -0.12}\) & \(20.9\substack{+1.8 \\ -1.7}\) \\
& \(0.726\substack{+0.024 \\ -0.027}\) & \(22.8\substack{+0.5 \\ -0.4}\) & \(5.2\substack{+0.5 \\ -0.5}\) & \(-7.6\substack{+0.4 \\ -0.5}\)\\
\midrule
\multirow{2}{*}{G2 \& SM}& & \(18.32\substack{+0.14 \\ -0.13}\) & \(10.1\substack{+0.4 \\ -0.4}\) \\
 & \(0.842\substack{+0.028 \\ -0.029}\) & \(40.1\substack{+0.9 \\ -0.7}\) & \(4.7\substack{+0.4 \\ -0.5}\) & \(-3.57\substack{+0.16 \\ -0.18}\)\\
\midrule
\multirow{2}{*}{Eilers et al.} & & \(8.43\substack{+0.15 \\ -0.13}\) & \(13.6\substack{+1.4 \\ -1.4}\) \\
& \(0.281\substack{+0.016 \\ -0.017}\) & \(19.4\substack{+0.3 \\ -0.3}\) & \(4.21\substack{+0.4 \\ -0.3}\) & \(-38.1\substack{+3.9 \\ -5.1}\)\\ 
\bottomrule
\end{tabular}
\end{table}

\begin{table}[t]
\caption{Results for the full data set (no cut-off). The table shows the BIC, Bayes ratio and $\chi^2\indice{red}$ values for all baryonic models.  For the differences between the BICs of the standard and modified gravity case, we report the absolute value.}
\label{tab:all_data_modelselection}
\centering
\begin{tabular}{llllll}
\toprule
baryonic model & & BIC & $\Delta \mathrm{BIC}$ & \(2 \log\indice{e} B_{12}\) \(\left(\frac{\beta \neq 0}{\beta =0}\right)\) & $\chi^2\indice{red}$\\
\midrule
\multirow{2}{*}{E2 \& TT} & stand. grav. & 7599.8 & \multirow{2}{*}{1808.9}& \multirow{2}{*}{1970.6} & 2.69\\
& mod. grav. & 5790.9 & & & 2.05\\
\midrule
\multirow{2}{*}{E2 \& SM} & stand. grav. & 8634.2 & \multirow{2}{*}{2973.3} & \multirow{2}{*}{3134.2} & 3.06 \\
& mod. grav. & 5660.9 & & & 2.00 \\
\midrule
\multirow{2}{*}{G2 \& TT} & stand. grav. & 13295 & \multirow{2}{*}{7642.6} & \multirow{2}{*}{7803.0} & 4.72\\
& mod. grav. & 5652.4 & & & 2.00\\
\midrule
\multirow{2}{*}{G2 \& SM} & stand. grav.& 12613 & \multirow{2}{*}{6977.2} & \multirow{2}{*}{7136.5} & 4.47\\
& mod. grav. & 5635.8 &  & & 1.99 \\
\midrule
\multirow{2}{*}{Eilers et al.} & stand. grav. & 10407 & \multirow{2}{*}{4634.9} & \multirow{2}{*}{4638.3} & 3.69\\
& mod.grav. & 5772.1 &  &  & 2.04\\
\bottomrule
\end{tabular}
\end{table}

\begin{table}[t]
\caption{Results for the full data set (no cut-off). The table shows the values of the Bayes ratio for the standard gravity case, which is used to distinguish between the different baryonic models. The best-fitting model is highlighted.}
\label{tab:bayesratio_stand_grav}
\centering

\begin{tabular}{c|c|cccc}
\multicolumn{1}{c}{} & \multirow{2}{*}{\(2 \log\indice{e} B_{12}\)} & \multicolumn{4}{c}{Model 1}\\
\cline{3-6}
\multicolumn{1}{c}{} &  & \textbf{E2 \& TT} & E2 \& SM & G2 \& TT & G2 \& SM \\
%\midrule
\hline
\parbox[t]{2mm}{\multirow{4}{*}{\rotatebox[origin=c]{90}{Model 2}}} & E2 \& TT & 0 & -1034.4 & -5695.4 & -5013.2\\
& E2 \& SM &  & 0 & -4661.0 & -3978.8\\
& G2 \& TT &  &  & 0 & 682.14\\
& G2 \& SM &  &  &  & 0 \\
& Eilers et al. & 2646.8 & 1611.6 & -3050.0 & -2368.8\\
\hline
\end{tabular}
\end{table}

\subsubsection{Modified gravity}
\begin{figure}[t]
\begin{subfigure}{.5\textwidth} 
\centering
\includegraphics[scale=0.6]{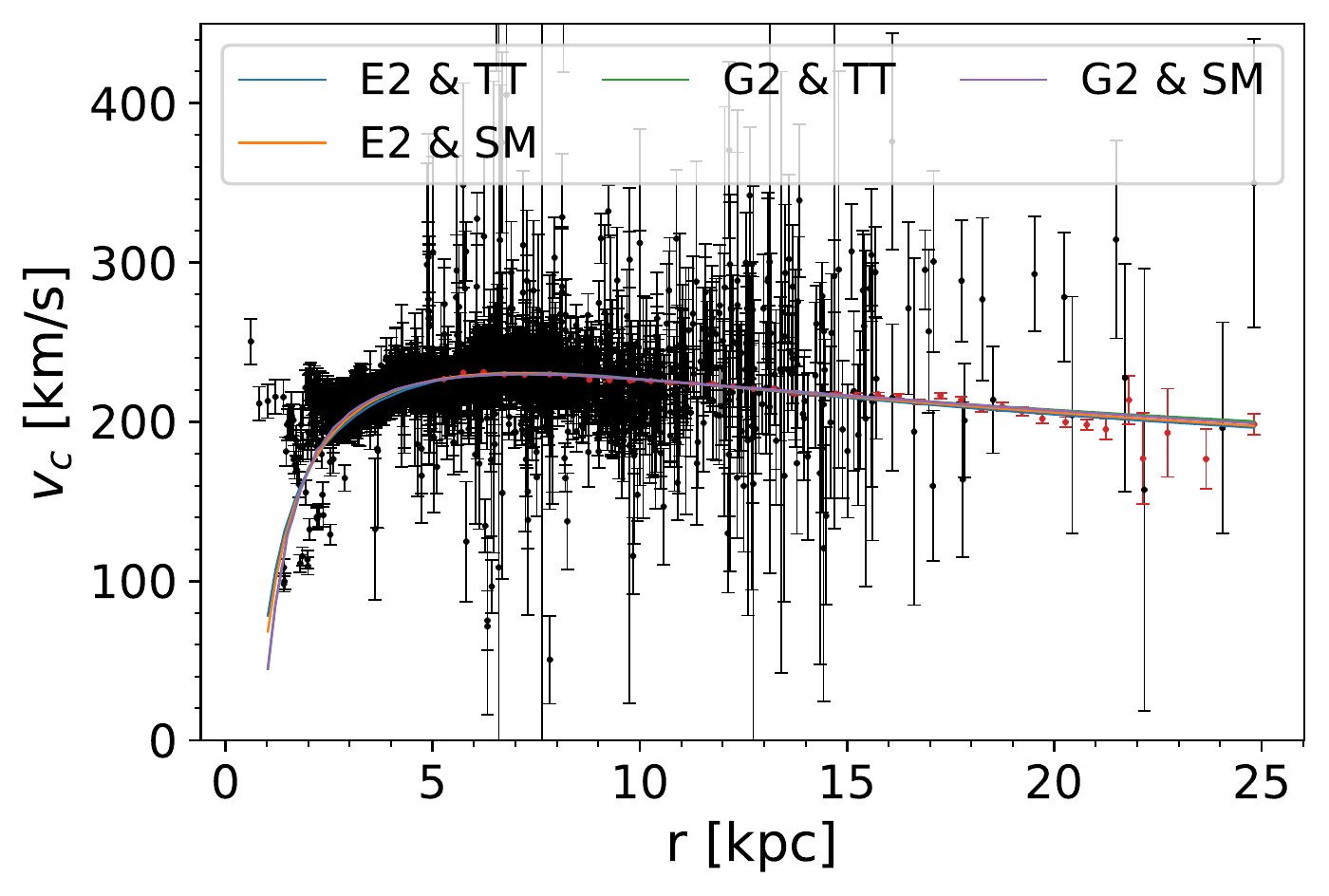}
\end{subfigure} 

\begin{subfigure}{.5\textwidth} 
\centering
\includegraphics[scale=0.6]{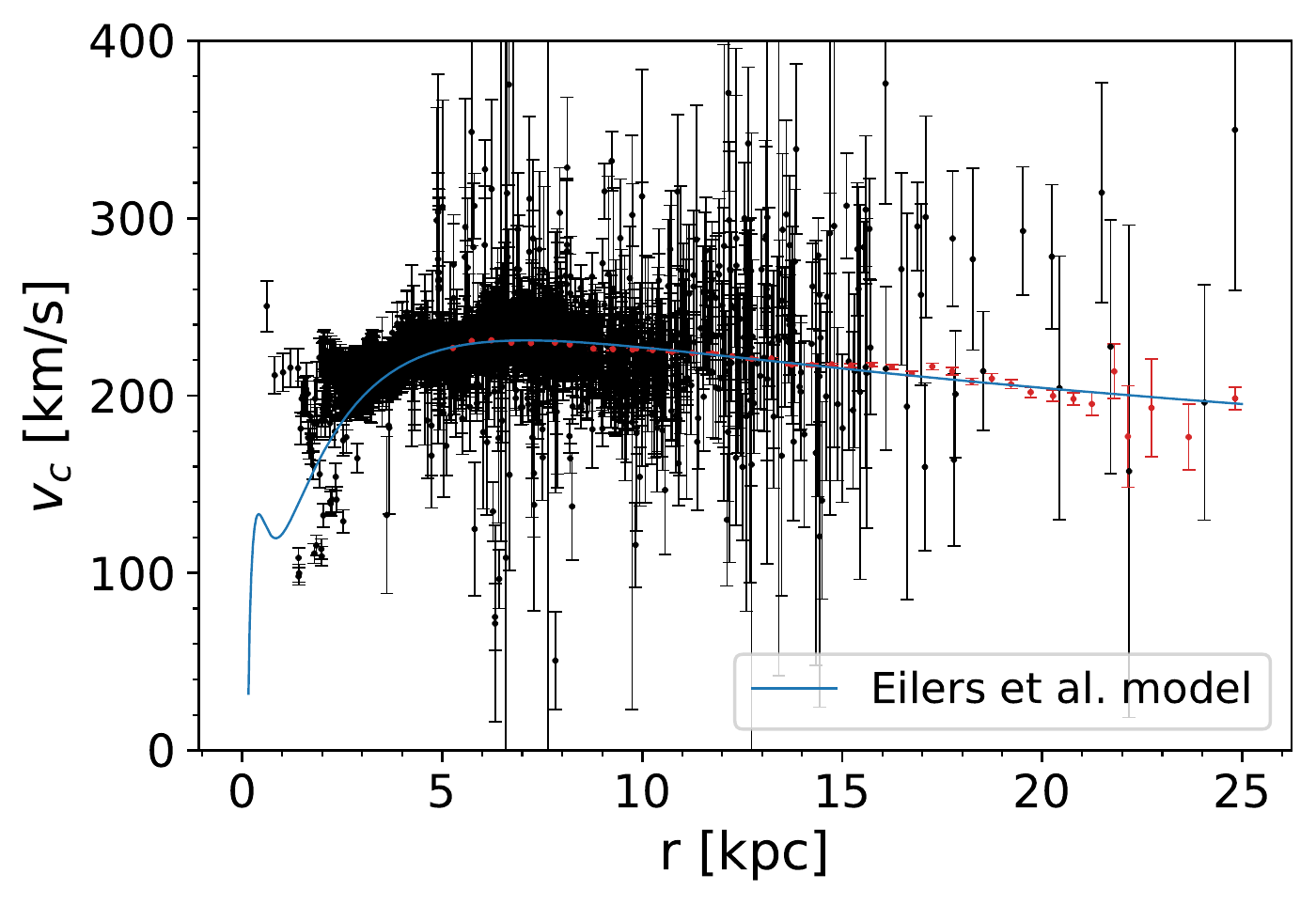}
\end{subfigure}
\begin{subfigure}{.5\textwidth} 
\centering
\includegraphics[scale=0.6]{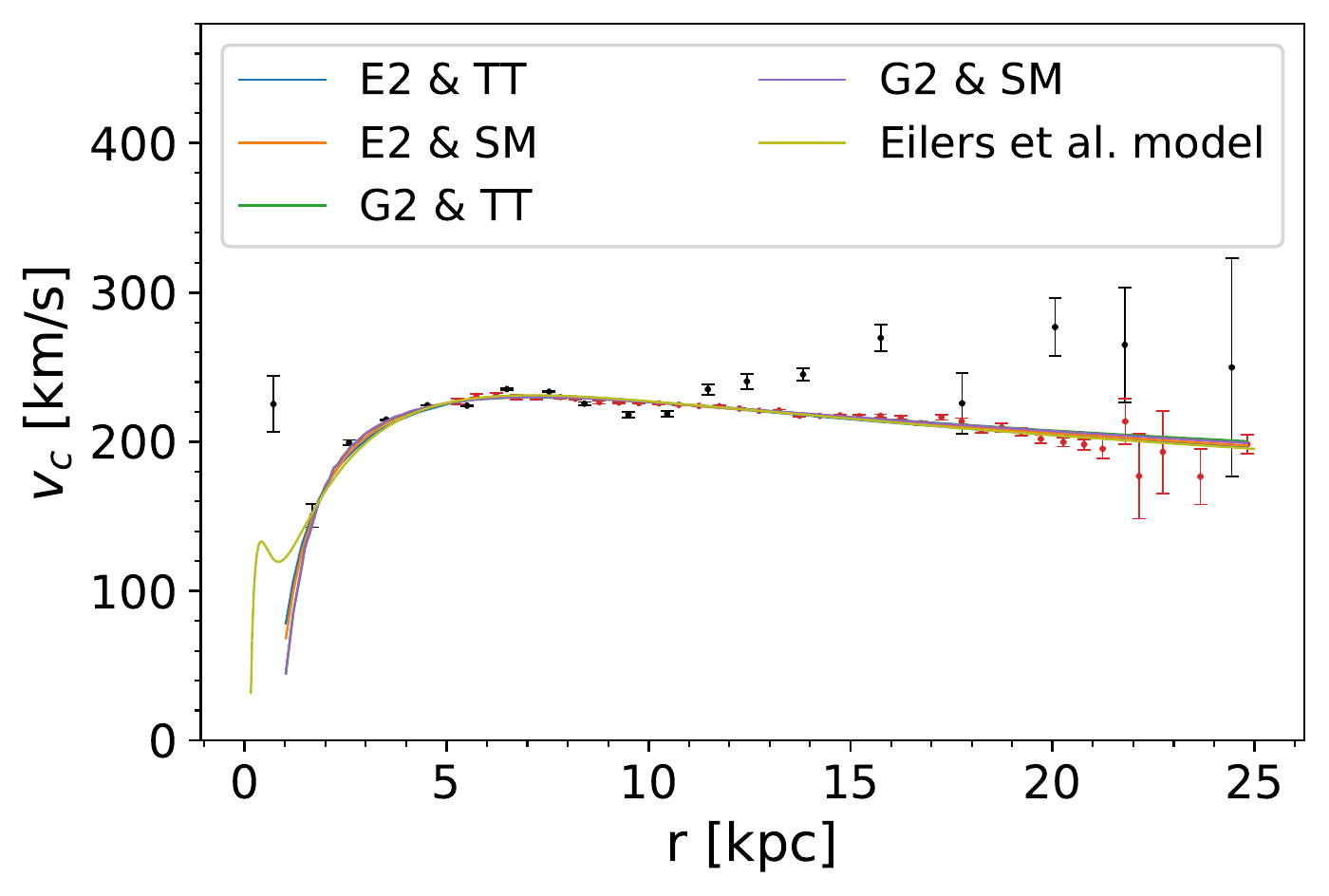}
\end{subfigure}
\caption{Modified gravity and full data set (no cut-off). The first plot shows the rotation curve fitted with our four baryonic models and the second plot shows the rotation curve fitted with the Eilers et al. baryonic model. The last plot shows the rotation curve for all 5 baryonic models with the data set {\it galkin} binned. The black data points are from the {\it galkin} data set and red data points are from \cite{Eilers2018}.}
\label{fig:rotation_curve_mod_grav_all_data}
\end{figure}

\begin{figure}[t]
\centering
\includegraphics[scale=0.6]{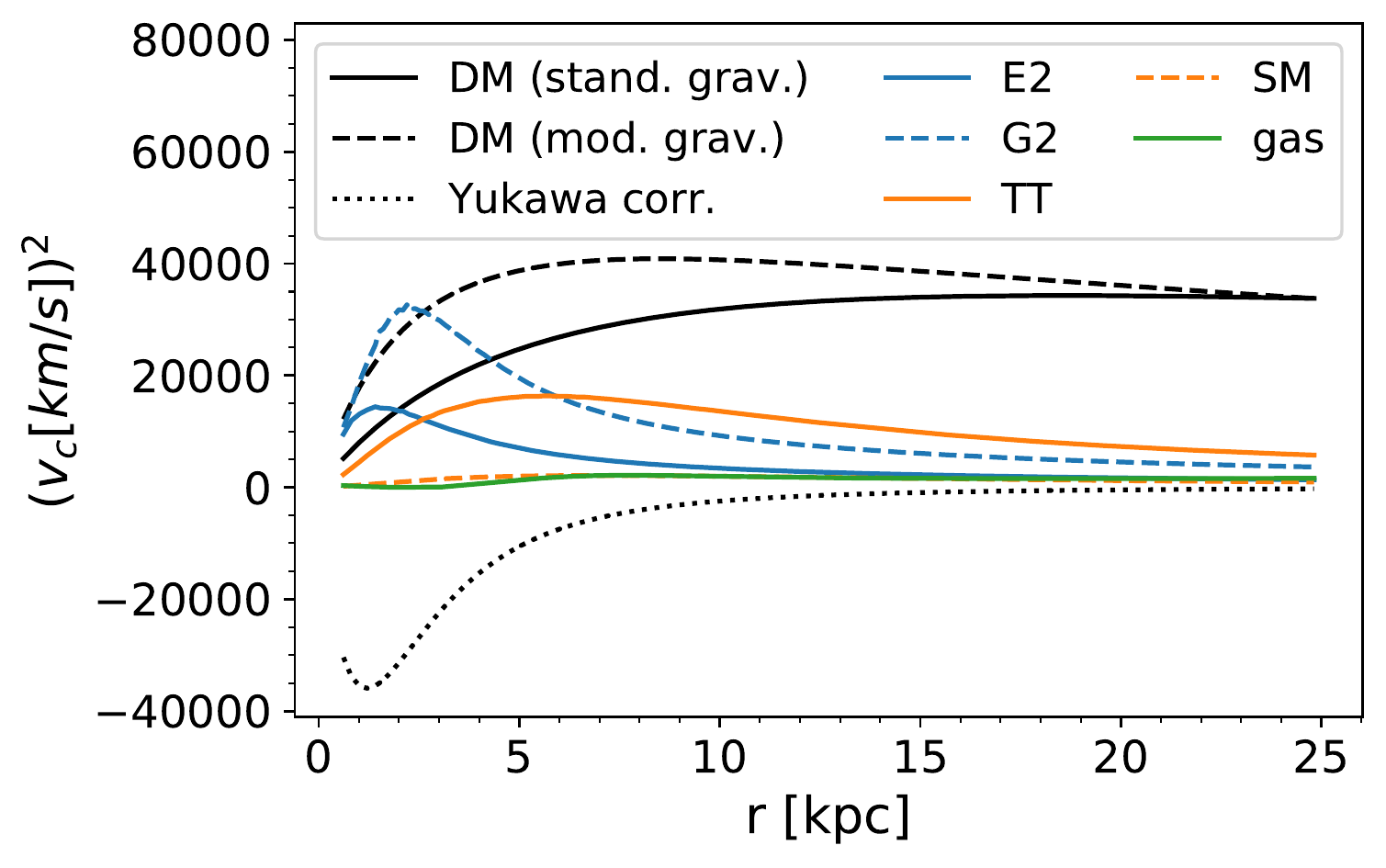}
\caption{Results for the full data set. This plot shows the contribution of all different components to the rotation curve. For the standard and modified gravity contribution, only the result for the best fitting case is shown.}
\label{fig:rotation_curve_components_fulldata}
\end{figure}

The best-fitting parameters for the modified gravity case are shown in Table~\ref{tab:results_all_data}, while the corresponding rotation curves are shown in Fig.~\ref{fig:rotation_curve_mod_grav_all_data}. The contribution of each component to the rotation curve can be seen in Fig.~\ref{fig:rotation_curve_components_fulldata}

As in the previous analyses, we find a repulsive Yukawa correction, meaning negative values for the the Yukawa strength $\beta$ for all baryonic models. However, $\beta$ is incompatible with zero this time; on the contrary, the relatively large value reduces the contribution of both the dark and the baryonic matter to the total rotation curve, resulting in a relative fast decline at small radii of the rotation curve, as it can be seen in Fig.~\ref{fig:rotation_curve_mod_grav_all_data}.
\\
Moreover, we find again small values for the the Yukawa length $\lambda$; in other words, the Yukawa correction is only dominant for small radii and reduces to $\sim (10^{-9} - 10^{-5})\cdot\beta$ at $R\simeq 8\, \mathrm{kpc}$ depending on the baryonic model.

We also write the NFW parameters in terms of the concentration parameter and the virial mass of the dark matter halo. The corresponding values are shown in Table~ \ref{tab:results_all_data_mc}. As before, the amount of dark matter needed to describe the rotation curve in the modified gravity case is not deviating significantly for the different baryonic models.
Again, the amount of dark matter needed for the modified gravity case is less, than the amount needed for the standard gravity case. Depending on the chosen baryonic model, the amount of dark matter needed is reduced by roughly $25\% - 75\%$.

The resulting ﬁt for the Eilers et al. baryonic model can be seen in Fig.~\ref{fig:rotation_curve_mod_grav_all_data}. To get this converging ﬁt, we have to extend the prior of the Yukawa strength in the negative range up to -100. 
The ﬁrst bump in the ﬁtted rotation curve should be due to the baryonic model used for the bulge, which has a very high peak for very small radii. 
As we compare these results with the results obtained for the other baryonic models, we immediately notice a very large value of the Yukawa strength.
This could be framed again in the context of the model used for the bulge. The bulge in the Eilers et al. has a very high contribution at small radii. Hence, a very strong repulsive Yukawa correction at small radii is needed to reduce the contribution of the dark matter and also of the baryonic matter to the total rotation curve.

For the model selection, we use as for the standard gravity case all three statistical indicators. The values for the $\chi^2\indice{red}$ and BIC are shown in Table~\ref{tab:all_data_modelselection} and the values of the Bayes ratio are shown in Table~\ref{tab:bayesratio_mod_grav}. 

The Bayes ratio shows that the model G2 \& SM is favoured against all the other baryonic models.  
Making use of the Kass-Raftery scale, we find a strong evidence in favour of the G2 \& SM model. This is also supported by the BIC values. 
The $\chi^2\indice{red}$ values are smaller than the values found in the standard gravity case; nevertheless, they are still quite large, showing that the rotation curve in Fig.~\ref{fig:rotation_curve_mod_grav_all_data} is a ``bad" fit.

To conclude, we compare the results obtained in standard gravity with the ones of modified gravity. The corresponding values for the Bayes ratio are shown in Table~\ref{tab:all_data_modelselection}. 
The values show a very strong evidence in favour of the modified gravity cases, for all the baryonic models considered. An explanation for such a strong preference, also confirmed looking at the differences between the BIC values, can be found in the fact that the Yukawa correction allows the rotation curve to decline fast enough at small radii. Roughly speaking, the modified gravity is able to better describe the data points in the region $r < 3\, \mathrm{kpc}$: enough large negative value of $\beta$ shall ensure a lower contribution of both the dark and the baryonic matter components to the total rotation curve.

We emphasize once more that the $r<3$ kpc region cannot be well modeled by circular rotation curves and therefore the conclusions of this section, reported for completeness and further analysis, cannot be taken as evidence for modified gravity.

\begin{table}[t]
\caption{Full data set (no cut-off). The table shows the values of the Bayes ratio for the modified gravity case, which is used to distinguish between the different baryonic models. The best-fitting model is highlighted.}
\label{tab:bayesratio_mod_grav}
\centering

\begin{tabular}{c|c|cccc}
\multicolumn{1}{c}{} & \multirow{2}{*}{\(2\log\indice{e} B_{12}\)} & \multicolumn{4}{c}{Model 1}\\
\cline{3-6}
\multicolumn{1}{c}{} &  & E2 \& TT & E2 \& SM & G2 \& TT & \textbf{G2 \& SM} \\
%\midrule
\hline
\parbox[t]{2mm}{\multirow{4}{*}{\rotatebox[origin=c]{90}{Model 2}}} & E2 \& TT & 0 & 129.2 & 137.0 & 152.7\\
& E2 \& SM &  & 0 & 7.837 & 23.53\\
& G2 \& TT &  &  & 0 & 15.69\\
& G2 \& SM &  &  &  & 0 \\
& Eilers et al. & -16.18 & 113.0 & 120.8 & 136.5 \\
\hline
\end{tabular}
\end{table}

%%%%%%%%%%%%%%%%%%%%%%%%%%%%%%%%%%%%%%%
%%%%%%%%%%%%%%%% Sec IX %%%%%%%%%%%%%%%
%%%%%%%%%%%%%%%%%%%%%%%%%%%%%%%%%%%%%%%
\section{Conclusions}\label{section:conclusion}

In this paper, we have used the Milky Way to search for constraints on gravity when a Yukawa term is included. 
We have adopted several data--sets for the Rotation Curve of our Galaxy, the first time that such comprehensive analysis is performed to reach this goal, at the best of our knowledge.

The originality of our analysis lies also in other respects: 1) we have included both dark matter and the Yukawa correction; 2) we let the coupling reach also negative values; 3) we have  modelled the baryonic sector in several alternative ways.
Moreover, we have analysed the results in terms of Bayesian model selection, giving to our results the proper statistical meaning.

The analysis is divided into three parts, each one with a different cut-off for the data set. In the case for the 3 kpc cut-off, we found that the Yukawa coupling is negative for all baryonic models and that the scaling length is very short ($\lambda < 1$ kpc). Using the BIC as a selection criteria, we found that the baryonic model E2 $\&$ TT for the standard gravity case fits the data best. For the analysis with a 5 kpc cut-off the conclusion is the same as for the 3 kpc cut-off: the baryonic model E2 $\&$ TT for the standard gravity case fits the data best. In the last part, with the complete data set, we get different results. Again, we found for the Yukawa coupling large negative values, but this time our model selection criteria prefer modified gravity over standard gravity. The most favoured baryonic model is G2 $\&$ SM.

\begin{figure}[t]
\begin{subfigure}{.5\textwidth} 
\centering
\includegraphics[scale=0.6]{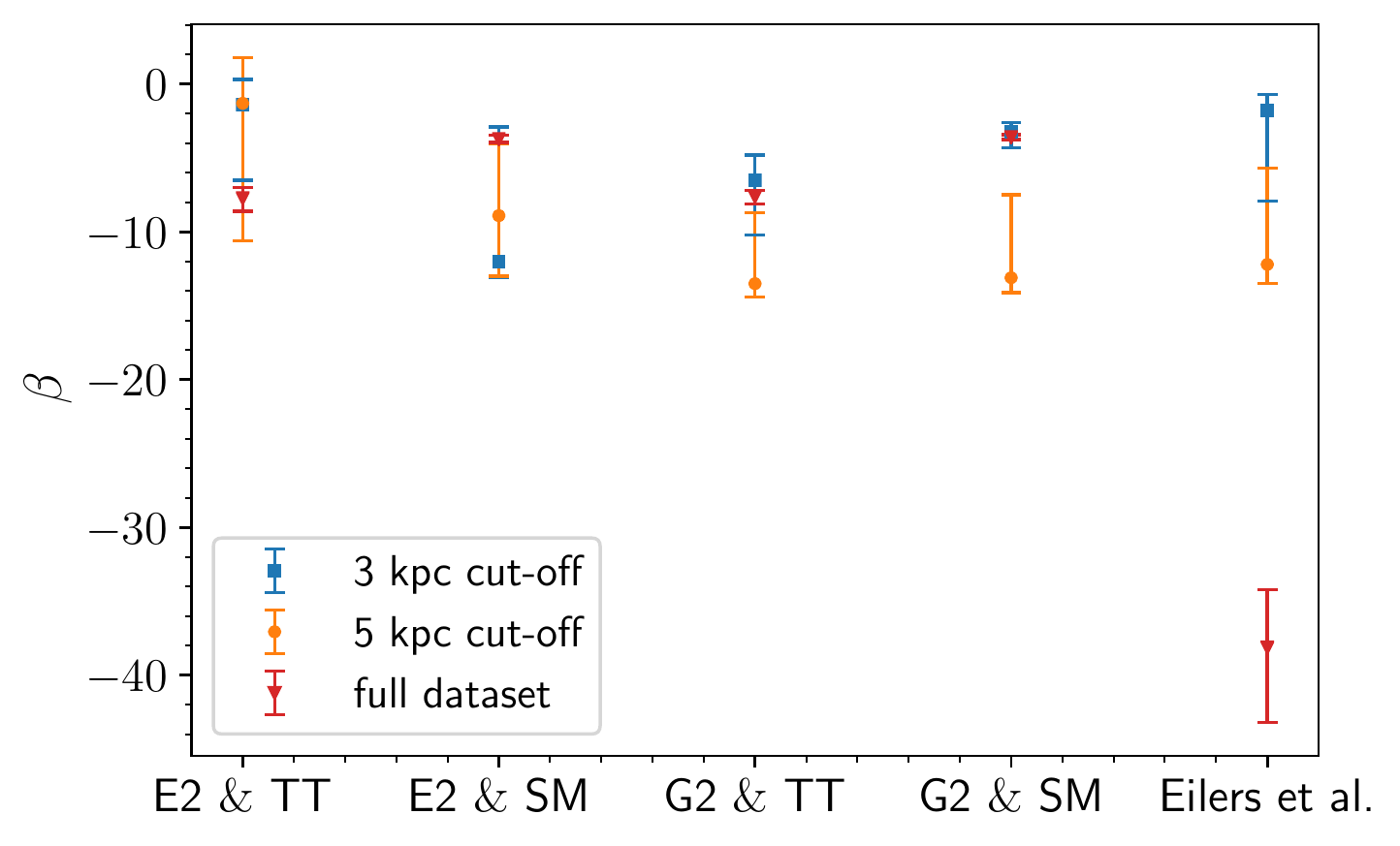}
\end{subfigure}
\begin{subfigure}{.5\textwidth} 
\centering
\includegraphics[scale=0.6]{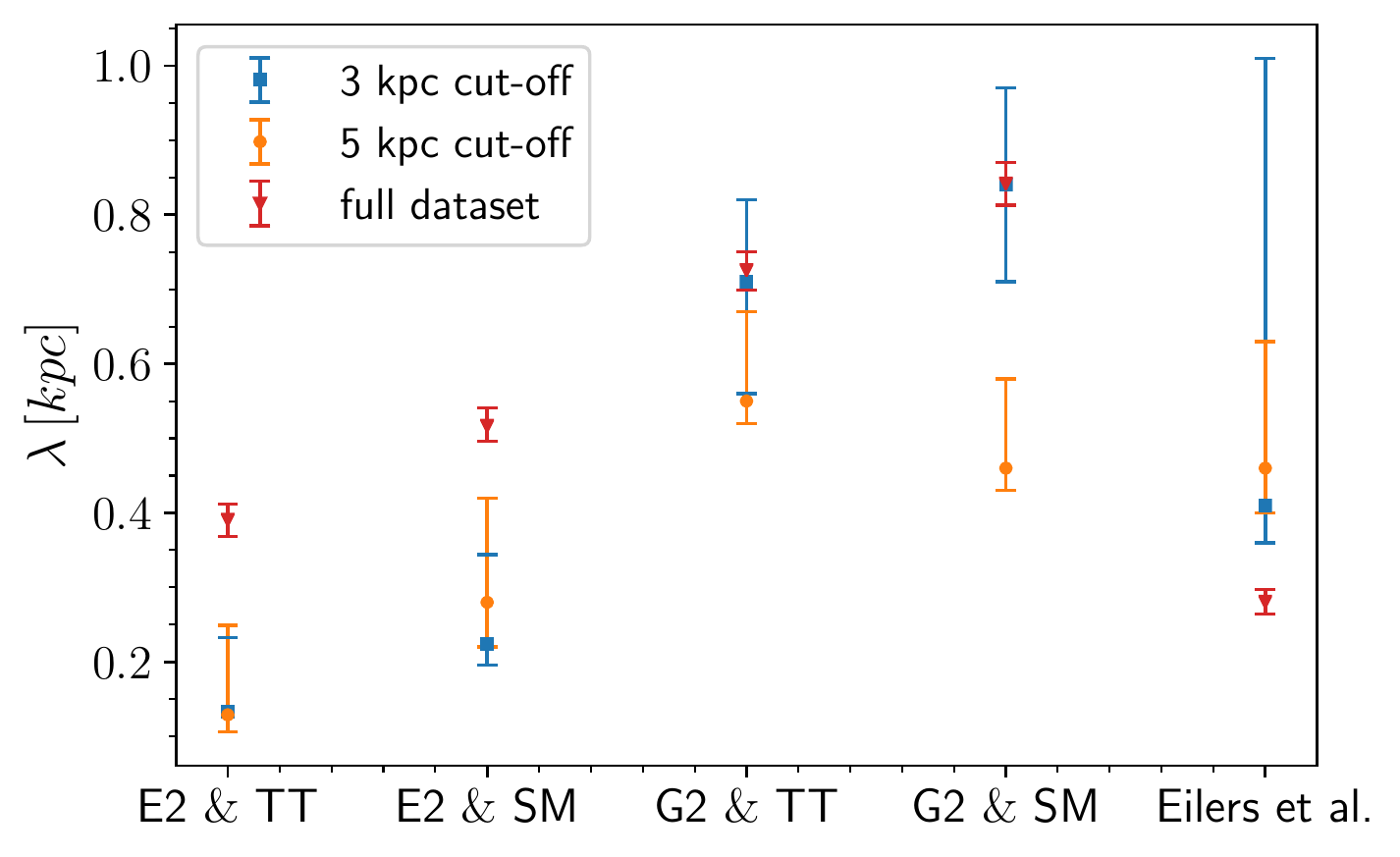}
\end{subfigure}
\caption{The plot shows a summary of the best fitting values for the Yukawa coupling and scaling length for the different baryonic models used in the analysis.}
\label{fig:summary_yukawa}
\end{figure}

The results are both intriguing and in line with expectations at the same time. We found a strong preference for a large and negative Yukawa term when we indiscriminately include all the rotation curve data. 
This conclusion would be very striking, were it not for the fact that the data at small scale themselves show an intrinsic spread quite larger than expected on a statistical level, thereby pointing to a systematic effect in the measurement. 
This suspect is quite substantiated by the analysis performed cutting off all the data below 3 kpc, where the data show much less intrinsic tension, and where the precise and homogeneous dataset by Eiler et al. give the dominant information. In this case, we found standard gravity as preferred model. 
The final constraint on the Yukawa coupling and scaling length using the best baryonic model for the 3 kpc cut-off are $\beta=-12.0^{+9.1}_{-1.1}$ and $\lambda=0.224^{+0.12}_{-0.028}\,\,\mathrm{kpc}$ (see Table~\ref{tab:results_cut_off_3}). The corresponding value for the field mass in eV is $m=\lambda^{-1}=2.85^{+0.41}_{-1.0}\times 10^{-26} \,\,\mathrm{eV}$. An overview of all best fitting Yukawa parameters ($\beta, \lambda$) can be found in Fig.~\ref{fig:summary_yukawa}.

The very large errors show that the $>3$ kpc dataset has a very weak constraining power on gravity. Moreover, there is a marked degeneracy between $\beta$  and $\lambda$ (see e.g. Fig.~\ref{fig:confidence_mod_grav_cutoff_3_E2SM}) so the individual constraints are not very informative. In Eq.~\eqref{eq:fitlambda} we present an approximate fit to the degeneracy curve.

One of the lessons of this work is that better baryonic models are crucial to improve the constraints on the rotation curve parameters, confirming in the context of modified gravity findings obtained in dark matter case only \cite{Iocco:2015iia,Iocco:2016itg}. The scatter between different models shown in Fig.~\ref{fig:summary_yukawa}, for instance, is clearly larger than statistically expected, especially for $\lambda$. Current GAIA data for tracers of the rotation curve do not help to sizably improve the status of things \cite{Benito:2020lgu}.

Although the final result 
is anything but unexpected, at least at large galactic distances, we believe that this kind of analysis is interesting and should be extended to further datasets, both in our Galaxy and in the other ones. 

\newpage
\section{Acknowledgements}
F.~I.’s work has been partially supported by the research grant number 2017W4HA7S “NAT-NET: Neutrino and Astroparticle Theory Network” under the program PRIN 2017 funded by the Italian Ministero dell’Università e della Ricerca (MUR).
%\printbibliography[title=References]
%\bibliography{bib/all_bib.bib}
\bibliography{all_bib.bib}
\end{document}